\documentclass[11pt, letterpaper]{article}

\usepackage[left=1in, right=1in, top=1in, bottom=1in]{geometry}
\usepackage[utf8]{inputenc}
\usepackage[T1]{fontenc}
\usepackage{bm}
\usepackage{type1cm}
\usepackage{lettrine}
\usepackage{amsmath,amssymb,amsthm}
\usepackage{moreverb}
\usepackage{mathtools}
\usepackage{amsmath}
\usepackage{amssymb}
\usepackage{algorithmic}
\usepackage{graphics}
\usepackage{graphicx}
\usepackage{subfigure}
\usepackage{caption}
\usepackage{extarrows}
\usepackage{color}
\usepackage{framed}
\usepackage{wrapfig}
\usepackage{bm}
\usepackage{mathrsfs}
\usepackage{mathabx}
\usepackage{multirow}
\usepackage{longtable}
\usepackage{hyperref}
\usepackage{paralist}
\usepackage{indentfirst}
\usepackage{relsize}
\usepackage{extarrows}
\usepackage{upgreek}
\usepackage{bm}
\usepackage{mwe}
\usepackage[dvipsnames,table]{xcolor}
\usepackage{booktabs}
\usepackage{authblk}
\usepackage{lettrine}
\usepackage{type1cm}
\usepackage{threeparttable}
\usepackage[sort&compress,numbers]{natbib}
\usepackage[figurename=Figure]{caption}
\usepackage[normalem]{ulem}

\graphicspath{ {./Figure/} }
\usepackage[font=footnotesize,labelfont=bf]{caption}

\providecommand{\keywords}[1]{\textbf{\textit{Keywords: }} #1}

\usepackage{booktabs, pifont}
\newcommand{\cmark}{\ding{51}} 
\newcommand{\xmark}{\ding{55}} 

\hypersetup{
bookmarks=true,
bookmarksopen=true,
bookmarksnumbered=true,
unicode=false,
pdftoolbar=true,
pdfmenubar=true,
pdffitwindow=false,
pdfstartview={FitH},
pdftitle={My title},
pdfauthor={Author},
pdfsubject={Subject},
pdfcreator={Creator},
pdfproducer={Producer},
pdfkeywords={keywords},
pdfnewwindow=true,
colorlinks=true,
linkcolor=blue,
citecolor=blue,
filecolor=blue,
urlcolor=blue
}

\begin{document}

\title{\textbf{Stable spectral neural operator for learning stiff PDE systems from limited data}}

\author[1]{Rui Zhang}
\author[1]{Han Wan}
\author[2]{Yang Liu}
\author[1,$^*$]{Hao Sun}

\affil[1]{\small Gaoling School of Artificial Intelligence, Renmin University of China, Beijing, China}
\affil[2]{\small School of Engineering Science, University of Chinese Academy of Sciences, Beijing, China \vspace{18pt}}
\affil[*]{Corresponding author}

\date{}

\maketitle

\normalsize

\vspace{-18pt} 
\begin{abstract}
\small
Accurate modeling of spatiotemporal dynamics is crucial to understanding complex phenomena across science and engineering. However, this task faces a fundamental challenge when the governing equations are unknown and observational data are sparse. System stiffness, the coupling of multiple time-scales, further exacerbates this problem and hinders long-term prediction. Existing methods fall short: purely data-driven methods demand massive datasets, whereas physics-aware approaches are constrained by their reliance on known equations and fine-grained time steps. To overcome these limitations, we introduce an equation-free learning framework, namely, the Stable Spectral Neural Operator (SSNO), for modeling stiff partial differential equation (PDE) systems based on limited data. Instead of encoding specific equation terms, SSNO embeds spectrally inspired structures in its architecture, yielding strong inductive biases for learning the underlying physics. It automatically learns local and global spatial interactions in the frequency domain, while handling system stiffness with a robust integrating factor time-stepping scheme. Demonstrated across multiple 2D and 3D benchmarks in Cartesian and spherical geometries, SSNO achieves prediction errors one to two orders of magnitude lower than leading models. Crucially, it shows remarkable data efficiency, requiring only very few (2--5) training trajectories for robust generalization to out-of-distribution conditions. This work offers a robust and generalizable approach to learning stiff spatiotemporal dynamics from limited data without explicit \textit{a priori} knowledge of PDE terms.
\end{abstract}


\vspace{12pt} 
\section*{Introduction}

Spatiotemporal partial differential equations (PDEs) are the mathematical language of the natural world, describing fundamental phenomena in fields ranging from fluid dynamics and climate science to materials science~\cite{lomax2002fundamentals,palmer2014climate,gao2025ai}. Accurate models are essential for mechanistic understanding, prediction, and control. However, in many scientific and engineering applications, a significant barrier exists: the governing equations are often unknown or incomplete, and measurements are sparse and expensive to obtain. These difficulties are exacerbated in stiff PDE systems, where widely separated timescales impose severe step size restrictions and compromise long-term stability~\cite{curtiss1952integration}. Stiffness is not a mere numerical inconvenience, but an intrinsic property of ubiquitous multiscale processes, including reaction--diffusion dynamics~\cite{halatek2018rethinking}, turbulent flows~\cite{wu2019efficient}, and phase-field evolution~\cite{allen1975coherent}, where reliable forecasts must simultaneously capture rapid transients and slowly evolving structures.

Classical time integrators highlight a stability-cost trade-off when modeling stiff dynamics. Explicit Euler and Runge--Kutta schemes are simple to implement but become unstable unless the time step \(\delta t\) is prohibitively small in stiff regimes~\cite{butcher1996runge}. Implicit methods enlarge the stability region but require solving large linear or nonlinear systems at every step~\cite{butcher1964implicit}. To further balance stability and efficiency, integrating factor schemes~\cite{cox2002exponential,krogstad2005generalized} advance the stiff linear dynamics analytically while treating nonlinear terms explicitly, alleviating step size constraints. However, all of these methods presuppose full knowledge of the governing equations, which are often unavailable or only partially specified in real-world applications.

In the equation-free regime, data-driven neural PDE solvers learn solution operators directly from data, bypassing fine-grained spatiotemporal discretization at inference. Representative methods include DeepONet~\citep{lu2021learning,kontolati2024learning,kopanivcakova2025deeponet} and the Fourier Neural Operator (FNO)~\citep{li2021fourier,tran2023factorized,zhang2024deciphering}, which map initial or boundary conditions and forcing to future states. Once trained, these models yield rapid predictions for new conditions, potentially accelerating the simulation of stiff PDE dynamics. Extensions based on transformers~\citep{wu2024transolver,li2023scalable} and graphs~\citep{pfaff2020learning,brandstetter2022message,zeng2025phympgn,mi2025conservation} improve scalability to complex domains and irregular meshes. Beyond deterministic solvers, generative models for spatiotemporal fields are explored for data assimilation, uncertainty quantification, and inpainting of missing measurements~\citep{du2024conditional,gao2024generative,li2024learning}. Nevertheless, purely data-driven approaches require large sets of high-fidelity trajectories, which are often obtainable only via costly experiments or large-scale simulations~\citep{parente2024data,li2024learning,zhang2025artificial}. Moreover, current data-driven solvers struggle under distribution shift, and out-of-distribution (OOD) robustness remains a central challenge~\citep{azizzadenesheli2024neural,zhang2025artificial}. 

To reduce data demand and improve robustness, physics-aware learning incorporates domain knowledge either in the loss (physics-informed) or in the architecture (physics-encoded)~\citep{faroughi2024physics}. Physics-informed approaches penalize PDE residuals (e.g., PINN~\citep{raissi2019physics}, PI-DeepONet~\citep{wang2021learning}) or enforce numerical-scheme consistency (e.g., PINO~\citep{li2024physics}, NeuralStagger~\cite{huang2023neuralstagger}, MCNP~\citep{zhang2025monte}) in the loss function. While effective when equations are known, they require explicit PDE forms and are sensitive to optimization settings~\citep{krishnapriyan2021characterizing}. By contrast, physics-encoded methods embed classical solvers into neural modules. In the finite-difference (FD) family, derivative stencils are instantiated as learnable convolutional kernels and paired with neural correctors to refine coarse solutions~\citep{long2019pde,rao2023encoding,ijcai2025p862,wang2025multipdenet}. The finite-volume (FV) family combines conservative flux updates with learned closures or subgrid terms~\citep{kochkov2021machine,yan2025learnable}. Both FD- and FV-based encodings inherit local receptive fields from stencil or flux updates. Consequently, they rely on hand-crafted modules to model global interactions, rather than learning the corresponding long-range operators end-to-end. Although recent spectral encoding methods extend long-range coupling in the frequency domain~\citep{dresdner2023learning,li2025symbolic}, these designs still rely on full PDE information or pre-defined PDE terms. Moreover, the time-stepping in current physics-encoded methods typically requires dense-in-time supervision (i.e., small $\delta t$), thereby increasing data-acquisition burden and slowing inference. This limitation is especially pronounced for stiff dynamics, where stability constraints and rapid transients force even smaller time steps, further increasing sampling requirements and runtime.

Despite recent advances, accurate long-term prediction of stiff PDE systems has yet to be achieved in an equation-free setting with limited and sparsely sampled data. Extended Data Table~\ref{tab:extendtab1} summarizes the strengths and limitations of representative approaches under this regime. To address these challenges, we introduce the \textbf{S}table \textbf{S}pectral \textbf{N}eural \textbf{O}perator (\textbf{SSNO}), a spectrally inspired neural solver that accommodates unknown physics and scarce data while delivering stable long-term rollouts in stiff regimes. Spatially, SSNO operates in the frequency domain to learn spatial derivatives and related operators without explicit PDE terms. Unlike FD- and FV-based methods, SSNO supports high-fidelity long-range interactions. Temporally, an integrating factor treatment of the stiff linear component combined with a high-order explicit update permits larger stable time steps and longer rollouts while avoiding large linear solves and maintaining stability across widely separated time scales. We further impose a sparsity prior on the learned operators to yield a compact, PDE-like structure, showcasing its interpretability. By embedding physical structure in both space and time, SSNO trains effectively from limited trajectories with large time steps and exhibits robust OOD generalization via strong inductive biases.

We validate SSNO across a diverse suite of 2D and 3D benchmarks, covering both Cartesian and spherical geometries, and including reaction-diffusion systems and incompressible Navier-Stokes equations. SSNO consistently surpasses strong data-driven and physics-encoded baselines in long-term predictive accuracy, achieving 1--2 orders of magnitude lower relative error and better spectral and statistical consistency. Even with only 2--5 training trajectories and no \textit{a priori} knowledge of explicit PDE terms, SSNO generalizes robustly to OOD settings and preserves fine-scale structure in stiff, multiscale regimes. Moreover, the sparsity regularizer used during training yields intermediate representations that align with the differential operators in PDE formulations, indicating physics-consistent structure rather than mere pattern memorization. Taken together, these results establish SSNO as a data-efficient, stable, and accurate neural solver for modeling and simulating stiff spatiotemporal dynamics.

\section*{Results}

\begin{figure}[!t]
   \centering
    \includegraphics[width=1.0\linewidth]{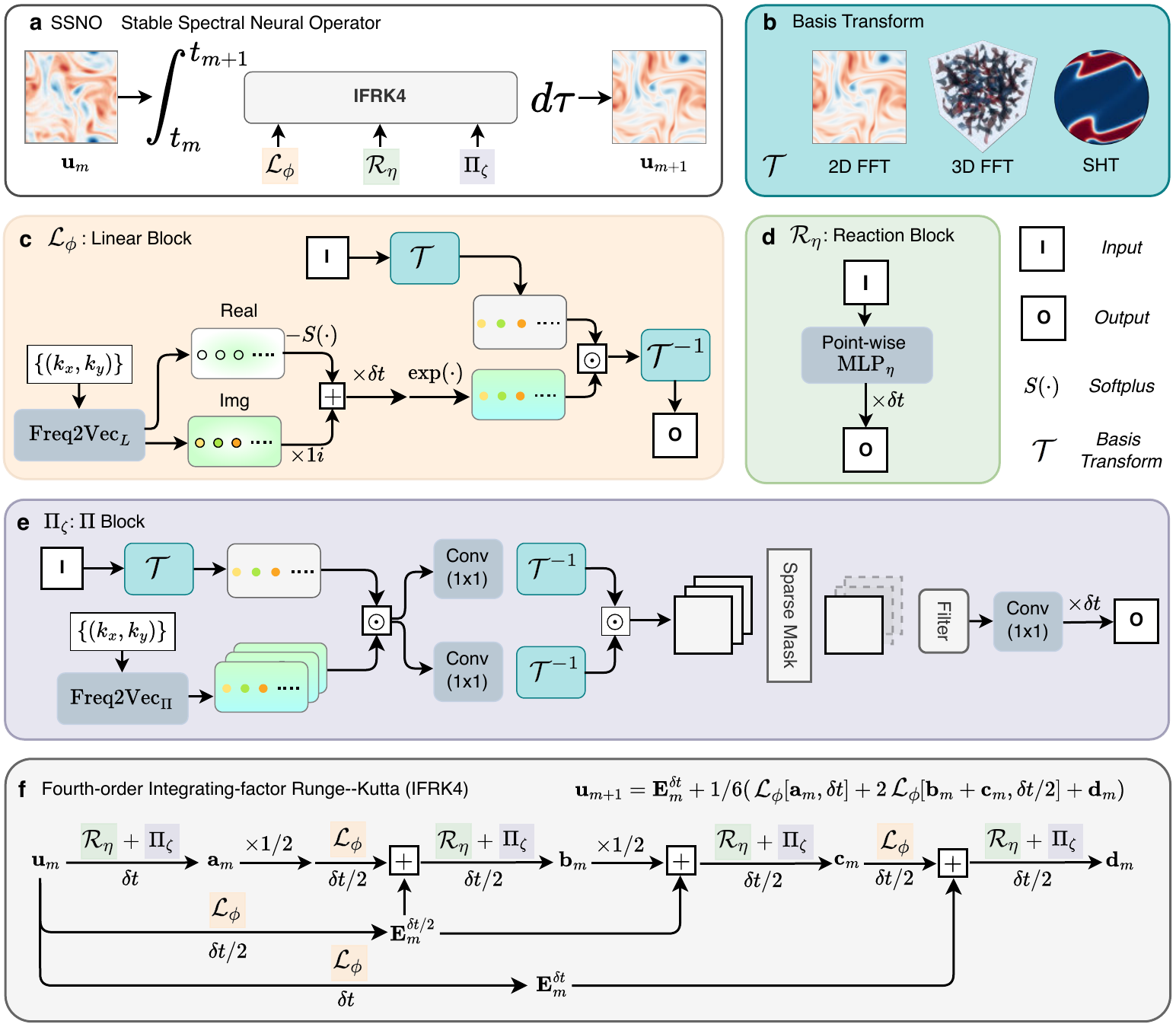}
\caption{\textbf{Architecture of the Stable Spectral Neural Operator (SSNO).}
\textbf{a}, A fourth-order integrating factor Runge--Kutta (IFRK4) step advances \(\mathbf{u}_m \!\to\! \mathbf{u}_{m+1}\) using three learnable parts: a linear block \(\mathcal{L}_{\phi}\), a local reaction term \(\mathcal{R}_\eta\), and a multiplication operator \(\Pi_{\zeta}\). \textbf{b}, Basis transforms \((\mathcal{T}, \mathcal{T}^{-1})\) map between physical and frequency domains via 2D/3D fast Fourier transforms (FFTs) or a spherical harmonic transform (SHT). \textbf{c--e}, Learnable blocks: \(\mathcal{L}_{\phi}\) is a frequency-wise operator to simulate linear operator; \(\mathcal{R}_\eta\) is a point-wise MLP in physical space to model reaction terms; \(\Pi_{\zeta}\) models nonlinear interactions via multiplication operators. \textbf{f}, The details of the IFRK4 scheme.}
\label{fig:1}
 \end{figure}

\subsection*{Overall framework of SSNO}

We consider spatiotemporal systems governed by the general form of PDEs as follows:
\begin{equation}
\frac{\partial \mathbf{u}(\mathbf{x},t)}{\partial t}
= \mathcal{L}[\mathbf{u}](\mathbf{x},t) + \Pi[\mathbf{u}](\mathbf{x},t) + \mathcal{R}[\mathbf{u}](\mathbf{x},t),
\quad \mathbf{x}\in\Omega.\label{eq:pde}
\end{equation}
where \(\mathbf{u}(\mathbf{x},t)\in \mathbb{R}^c\) denotes the physical field with $c$ channels, and $\mathbf{x}\in\mathbb{R}^d$ and $t$ denote the spatial and temporal coordinates, respectively. Here, $d$ represents the spatial dimension. The differential operator \(\mathcal{L}\) captures stiff linear dynamics (e.g., the diffusion term \(\nu\,\Delta \mathbf{u}\) and bi-Laplacian $\Delta^2 \mathbf{u}$). The operator \(\Pi\) represents multiplicative couplings, such as the convection term \(\mathbf{u}\cdot \boldsymbol{\nabla}  \mathbf{u}\) or gradient magnitudes \(\|\boldsymbol{\nabla}  \mathbf{u}\|_2^2\), where $\boldsymbol{\nabla}\in \mathbb{R}^d$ denotes the spatial gradient operator and $\Delta=\boldsymbol{\nabla}^2$ the Laplacian operator. \(\mathcal{R}\) denotes local reactions or sources, like the Allen--Cahn nonlinearity \(f(\mathbf{u})=\mathbf{u}-\mathbf{u}^3\). This framework in Eq.~\ref{eq:pde} abstracts a broad class of spatiotemporal systems from turbulent flows to reaction--diffusion processes. In this paper, we address a realistic yet challenging setting, where the governing PDEs in Eq.~\ref{eq:pde} are unknown and observations are simultaneously scarce (e.g., 2-5 training trajectories) and sparsely sampled in time (i.e., with a large time step $\delta t$ between observed snapshots). Our goal is to learn a neural PDE surrogate capable of autoregressive long-term prediction from limited data without requiring explicit PDE terms.

To this end, we propose the \textbf{S}table \textbf{S}pectral \textbf{N}eural \textbf{O}perator (\textbf{SSNO}), designed to balance stability, accuracy, and data efficiency for modeling stiff PDE systems in an equation-free manner (Fig.~\ref{fig:1}). Motivated by the structure of spectral methods, SSNO handles spatial operators primarily in the frequency domain via a transform pair \((\mathcal{T},\mathcal{T}^{-1})\). We use the fast Fourier transform (FFT) for periodic Cartesian 2D/3D grids and the spherical harmonic transform (SHT) for spherical geometry (Fig.~\ref{fig:1}\textbf{b}). In contrast to FD- or FV-encoded methods~\cite{rao2023encoding,yan2025learnable}, computing spatial derivatives in frequency space affords high-order accuracy with reduced discretization error and inherently models globally coupled physical interactions, rather than being constrained to local receptive fields~\cite{JML1268,shen2011spectral}.

Within the frequency domain, a lightweight Frequency-to-Vector (Freq2Vec) module maps frequency indices to corresponding multipliers, allowing the network to learn derivative-like symbols directly from data. The linear block \(\mathcal{L}_{\phi}\) advances the learned spatial-derivative operator via an integrating factor formulation, mitigating stiffness and relaxing step size constraints (Fig.~\ref{fig:1}\textbf{c}). We model nonlinearity through two specialized components: a point-wise reaction block \(\mathcal{R}_{\eta}\) in physical space (Fig.~\ref{fig:1}\textbf{d}), and a multiplicative block \(\Pi_{\zeta}\) that approximates products such as \(\mathbf{u}\cdot \boldsymbol{\nabla}  \mathbf{u}\) via learned compositions (Fig.~\ref{fig:1}\textbf{e}). To promote a compact, PDE-like structure, we impose an \(\ell_1\) penalty on channel combinations within \(\Pi_{\zeta}\) and apply a low-pass filter in frequency space to reduce aliasing before inverse transforms.

The learned linear and nonlinear blocks are then dynamically coupled within a fourth-order integrating factor Runge--Kutta (IFRK4) scheme (Fig.~\ref{fig:1}\textbf{f}). This staging permits larger stable time steps, supporting fast inference and reliable long-term rollouts under widely separated timescales.

In summary, SSNO represents a new class of neural operator where physical principles are not encoded via specific hard-crafted equation terms, but learned via the spectrally inspired structures to achieve strong inductive biases. This structure-encoded design, combining spectral operators with integrating factor time-stepping, is the key to achieving globally consistent accuracy from limited data, even at large time steps where previous methods fail. More details are given in the \textcolor{blue}{Methods} section.

To assess performance across diverse and challenging regimes, we assemble a benchmark suite of four representative PDEs spanning spatiotemporal chaos, pattern formation, turbulence, and phase separation on curved manifolds. Specifically, we consider (i) the Kuramoto--Sivashinsky equation (KSE) for its high-order derivatives and spatiotemporal chaos, testing numerical stability; (ii) the Swift--Hohenberg equation (SHE) for its characteristic pattern formation, testing fidelity in capturing emergent structures; (iii) the incompressible Navier--Stokes equations (NSE) for their non-local velocity-vorticity coupling, testing the necessity of a global representation; and (iv) the Allen--Cahn equation (ACE) on the sphere for its non-Euclidean geometry, testing OOD generalization to new manifolds. For all Euclidean problems, we impose periodic boundary conditions and generate data using high-fidelity spectral solvers. For the spherical problem, we employ a spherical-harmonic discretization with the differentiable \texttt{torch-harmonics} library~\citep{bonev2023spherical}. Detailed domains, resolutions, and time steps are provided in \textcolor{blue}{Supplementary Note~A}.

\subsection*{2D reaction--diffusion systems}

We consider two canonical 2D reaction--diffusion benchmarks with pronounced fourth-order stiffness: the Kuramoto--Sivashinsky equation (KSE) and the Swift--Hohenberg equation (SHE). The KSE, originally derived for laminar flame-front instabilities~\cite{sivashinsky1980flame} and widely used as a model of spatiotemporal chaos~\cite{michelson1977nonlinear}, in 2D takes the form
\begin{equation}
\label{eq:kse}
\frac{\partial u}{\partial t} + \Delta u + \Delta^2 u + \frac{1}{2}\,\|\boldsymbol{\nabla}  u\|_2^2 = 0.
\end{equation}
The competition between the anti-diffusive Laplacian ($-\Delta u$) and the stabilizing hyper-diffusive bi-Laplacian ($-\Delta^2 u$) gives rise to irregular spatiotemporal dynamics with long-wavelength modulations. The SHE models pattern-forming instabilities arising from finite-wavelength bifurcations~\cite{peletier2004pattern},
\begin{equation}
\label{eq:she}
\frac{\partial u}{\partial t} = u - (1+\Delta)^2 u - u^3,
\end{equation}
which selects a characteristic length scale and generates self-organized periodic structures. In both cases, the fourth-order spatial operator induces pronounced stiffness, providing a stringent test of stability and accuracy.

We benchmark SSNO against strong neural PDE baselines: Fourier-operator models DPOT~\cite{hao2024dpot} and FFNO~\cite{tran2023factorized}, the convolutional solver CNext~\cite{liu2022convnet,ohana2024well}, the transformer-based FactFormer~\cite{li2023scalable}, and the physics-encoded PeRCNN~\cite{rao2023encoding} with FD stencils embedded in its architecture. To stress learnability and stability under scarce, sparsely sampled data, we train with only five trajectories and roll out autoregressively on a \(64\times 64\) grid over \([0,8\pi)^2\) up to \(T=5\) with a time step \(\delta t=0.5\). This protocol evaluates whether models can maintain stable long-horizon prediction and preserve fine-scale structure in the presence of fourth-order stiffness.

\begin{figure}[!t]
   \centering
    \includegraphics[width=1.0\linewidth]{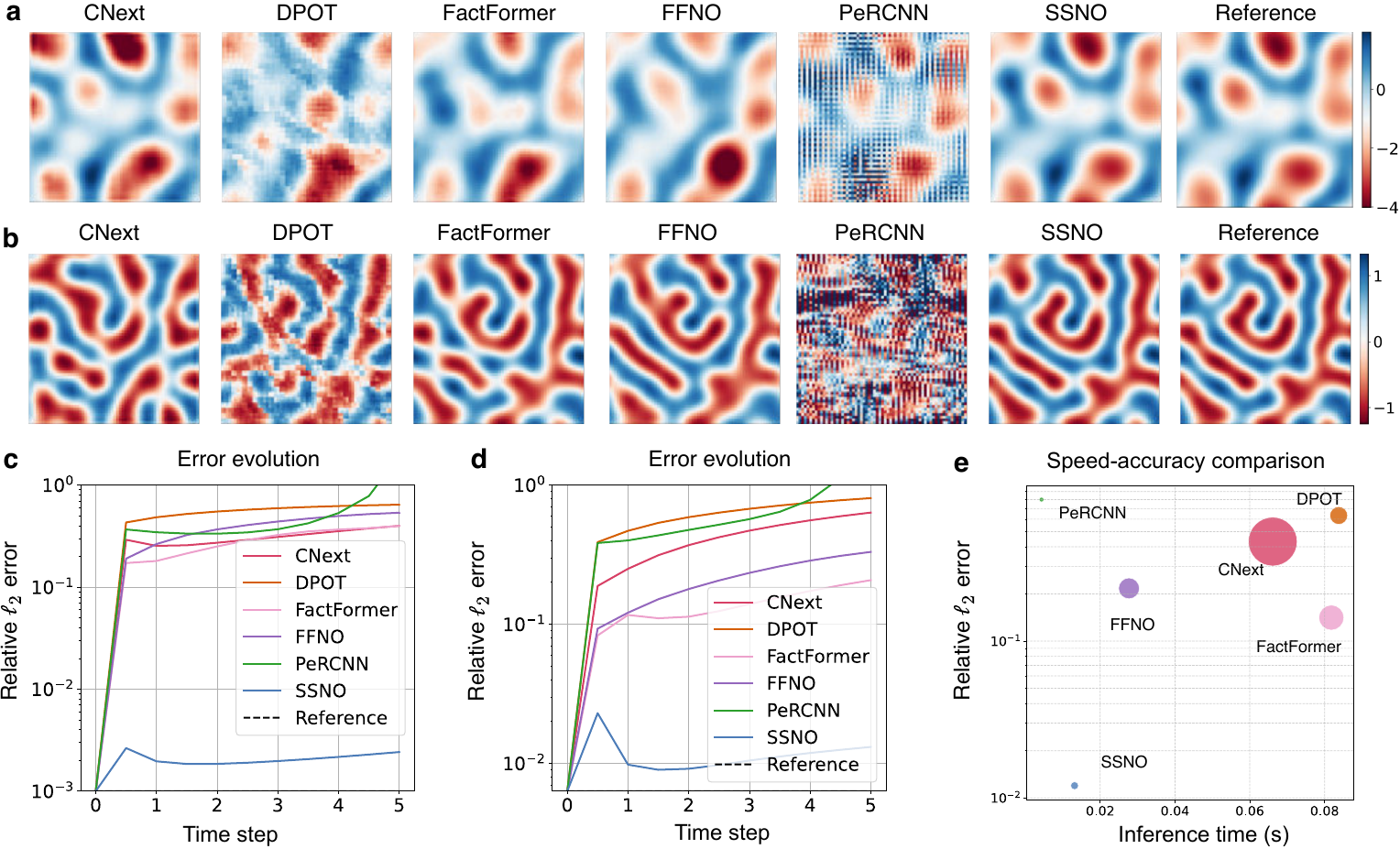}
\caption{\textbf{Multi-step error and predicted snapshots from SSNO and baselines on 2D reaction--diffusion systems.} \textbf{a} and \textbf{b}, Predicted fields for KSE and SHE at $t=5$. \textbf{c} and \textbf{d}, Multi-step relative $\ell_2$ error for KSE and SHE. \textbf{e}, Speed-accuracy comparison for SHE, where marker size represents the number of parameters of each model. The predicted evolutions for KSE and SHE are further depicted in \textcolor{blue}{Supplementary Figs. S.1--S.2}.}
\label{fig:2}
 \end{figure}

SSNO accurately reconstructs the reference phase and amplitude for both systems, preserving fine-grained features such as stripes and textures, and achieves substantially lower long-term relative $\ell_2$ error than all baselines (Fig.~\ref{fig:2}). By contrast, under this setting DPOT exhibits blocky artifacts and phase drift. Although FFNO, CNext, and FactFormer capture large-scale patterns, they struggle to generalize fine-scale structure with limited training data. PeRCNN diverges around \(t\!\approx\!4\), suggesting that its FD discretization and time marching are insufficiently stable for fourth-order stiffness at this time step (Fig.~\ref{fig:2}\textbf{a}--\textbf{b}). We also observe that SSNO's error exhibits an initial transient increase followed by a decline over time (Fig.~\ref{fig:2}\textbf{c}--\textbf{d}), consistent with fast early-stage dynamics that amplify small phase offsets; as the dynamics slow toward a quasi-steady regime, the prediction task becomes less demanding.

SSNO's advantage stems from two ingredients: accurate spectral differentiation and stable time-stepping. First, evaluating fourth-order operators in the frequency domain yields low-dissipation, high-order accuracy, reducing discretization error relative to FD-based physics-encoded methods. Second, the IFRK4 update propagates the stiff linear part analytically while treating nonlinear interactions explicitly, enabling robust rollouts even at a large time step of \(\delta t=0.5\). Robustness with respect to the rollout time step is further assessed in Extended Data Table~\ref{tab:extendtab2}: for 2D KSE, SSNO maintains very low relative error (on the order of $10^{-3}$) and near-perfect correlation ($>0.9998$) as \(\delta t\) increases tenfold (from 0.1 to 1.0), demonstrating strong stability.

A speed-accuracy comparison (Fig.~\ref{fig:2}\textbf{e}) indicates that SSNO attains higher accuracy at lower inference cost than other data-driven methods, while PeRCNN, although fast, is inaccurate for reliable multi-step prediction. Overall, with only five training trajectories and coarse time sampling on stiff reaction--diffusion systems, SSNO combines stability, accuracy, and efficiency.

\subsection*{2D incompressible Navier--Stokes equation}

To probe global coupling and long-term extrapolation, we consider the 2D vorticity equation
\begin{equation}
\begin{aligned}
    \frac{\partial \omega}{\partial t} + \mathbf{u}\cdot\boldsymbol{\nabla}  \omega 
    &= \frac{1}{\mathrm{Re}}\Delta \omega + f, \\
\boldsymbol{\nabla} \cdot\mathbf{u} &= 0, \\
\omega &= \boldsymbol{\nabla} \times\mathbf{u}, \label{eq:nse_vort}
\end{aligned}
\end{equation}
with a stationary Kolmogorov-type vorticity forcing (the curl of a sinusoidal body force), e.g. \(f(\mathbf{x})=0.1\sin(8x_1)\). We vary \(\mathrm{Re}\in\{500,1000,1500\}\) to span laminar to turbulent regimes. The velocity is recovered via a streamfunction formulation, solving \(\Delta \psi=\omega\) and setting \(\mathbf{u}=\boldsymbol{\nabla} ^\perp\psi\)~\cite{Saffman_1993}. Crucially, the recovery of the velocity field from vorticity requires solving a Poisson equation, which corresponds to an inherently non-local operation. This problem thus directly probes the advantage of SSNO's global spectral representation over methods confined to local FD- or FV-based stencils. 

We evaluate SSNO over the domain $[0,2\pi)^2$, discretized on a $128\times128$ grid. To assess long-term extrapolation, the training set comprises five trajectories truncated at $t=8$, while extending the testing time to $t=15$ with a fixed step size $\delta t=0.05$. The same set of baseline models from the 2D reaction--diffusion experiments is used for comparison.

\begin{figure}[!t]
    \centering
    \includegraphics[width=1\linewidth]{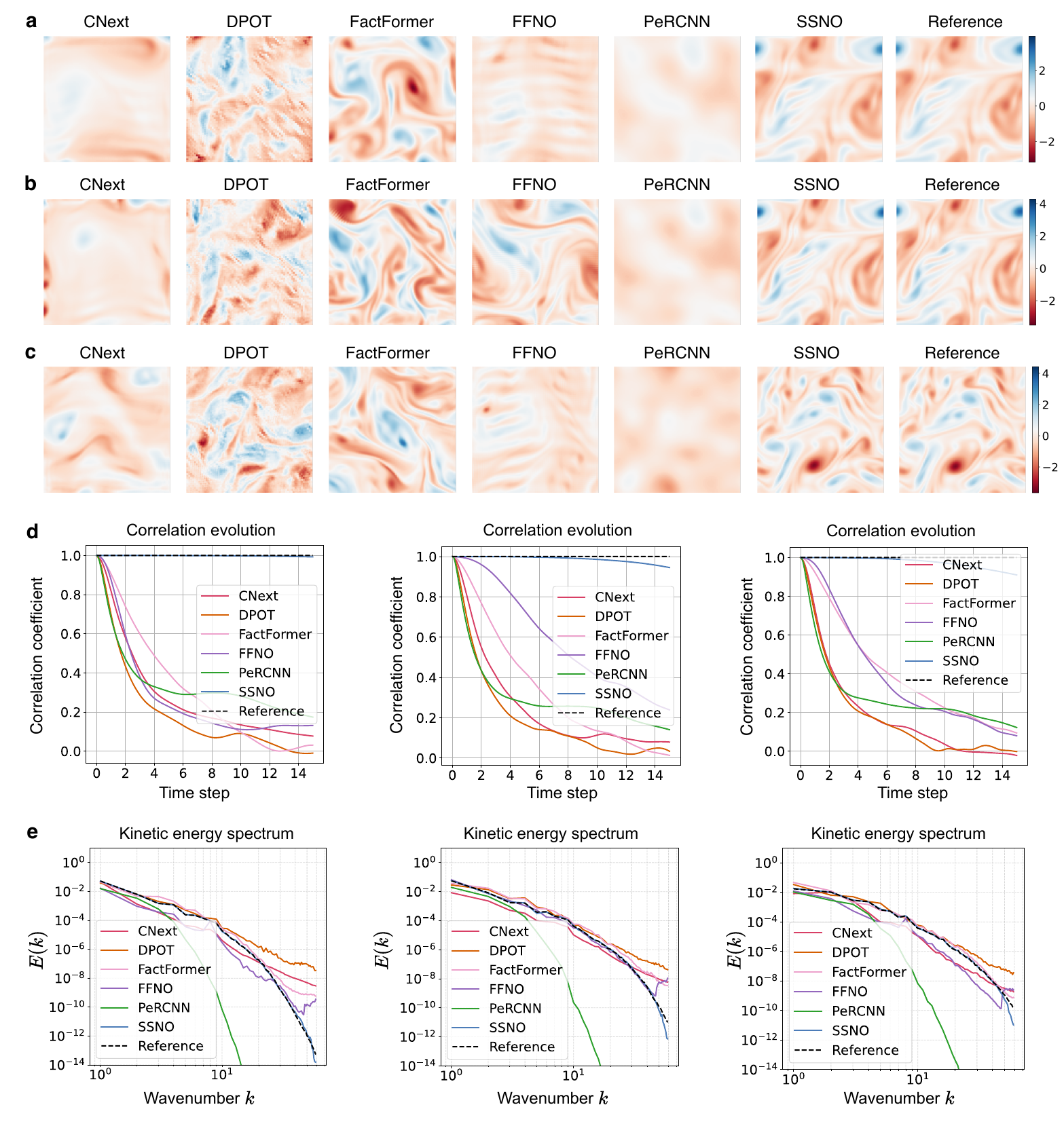}
\caption{\textbf{Multi-step error and predicted snapshots from SSNO and baselines on 2D NSE systems.} \textbf{a-c}, Predicted vorticity fields for NSE at $t=15$ for $\mathrm{Re}\in\{500,1000,1500\}$. \textbf{d}, Temporal evolution of correlation coefficients between predicted and reference vorticity fields for $\mathrm{Re}\in\{500,1000,1500\}$. \textbf{e}, Kinetic energy spectra of the predicted and reference vorticity fields for $\mathrm{Re}\in\{500,1000,1500\}$. The predicted flow evolutions are further depicted in \textcolor{blue}{Supplementary Figs. S.3--S.5}.}
    \label{fig:3}
\end{figure}

Fig.~\ref{fig:3}\textbf{a--c} show vorticity snapshots at $t=15$ for $\mathrm{Re}\in\{500,1000,1500\}$. SSNO's predictions remain in close agreement with the reference vorticity fields across all test cases. Even at $\mathrm{Re}=1500$, where nonlinearity is strongest, SSNO still preserves coherent vortical structures and fine-scale filaments. Quantitatively, the final-time correlation with the reference exceeds $0.9$ for all three Reynolds numbers (Fig.~\ref{fig:3}\textbf{d}). In contrast, the other models fail to produce meaningful predictions, as their correlations with the reference drop below 0.8 within the first four time steps. Notably, FactFormer still generates flow-like textures at $t=15$, yet the correlation with the reference remains low, indicating memorization rather than genuine extrapolation. PeRCNN's outputs bear little resemblance to the true flow. This behavior is consistent with the locality of FD stencils, which struggle to capture the globally coupled dynamics of the NSE. In contrast, SSNO's Freq2Vec module learns spectral multipliers that effectively model the global velocity-vorticity coupling (Eq.~\ref{eq:nse_vort}). Consequently, SSNO offers stronger global expressivity and higher numerical fidelity than FD-based physics-encoded models.

Furthermore, the energy spectra in Fig.~\ref{fig:3}\textbf{e} show that SSNO closely matches the reference across a broad range of wavenumbers, with only a slight deviation at the highest wavenumbers, whereas other methods exhibit a premature spectral roll-off. This discrepancy reflects an adaptive, data-driven spectral choice learned by SSNO during training: to sustain stable long-term rollouts, the model selectively attenuates the highest modes via the Freq2Vec module and low-pass masking, thereby preventing error amplification and numerical blow-ups during long-term rollouts.

\subsection*{Allen--Cahn on the sphere}
\begin{figure}[!t]
    \centering
    \includegraphics[width=0.92\linewidth]{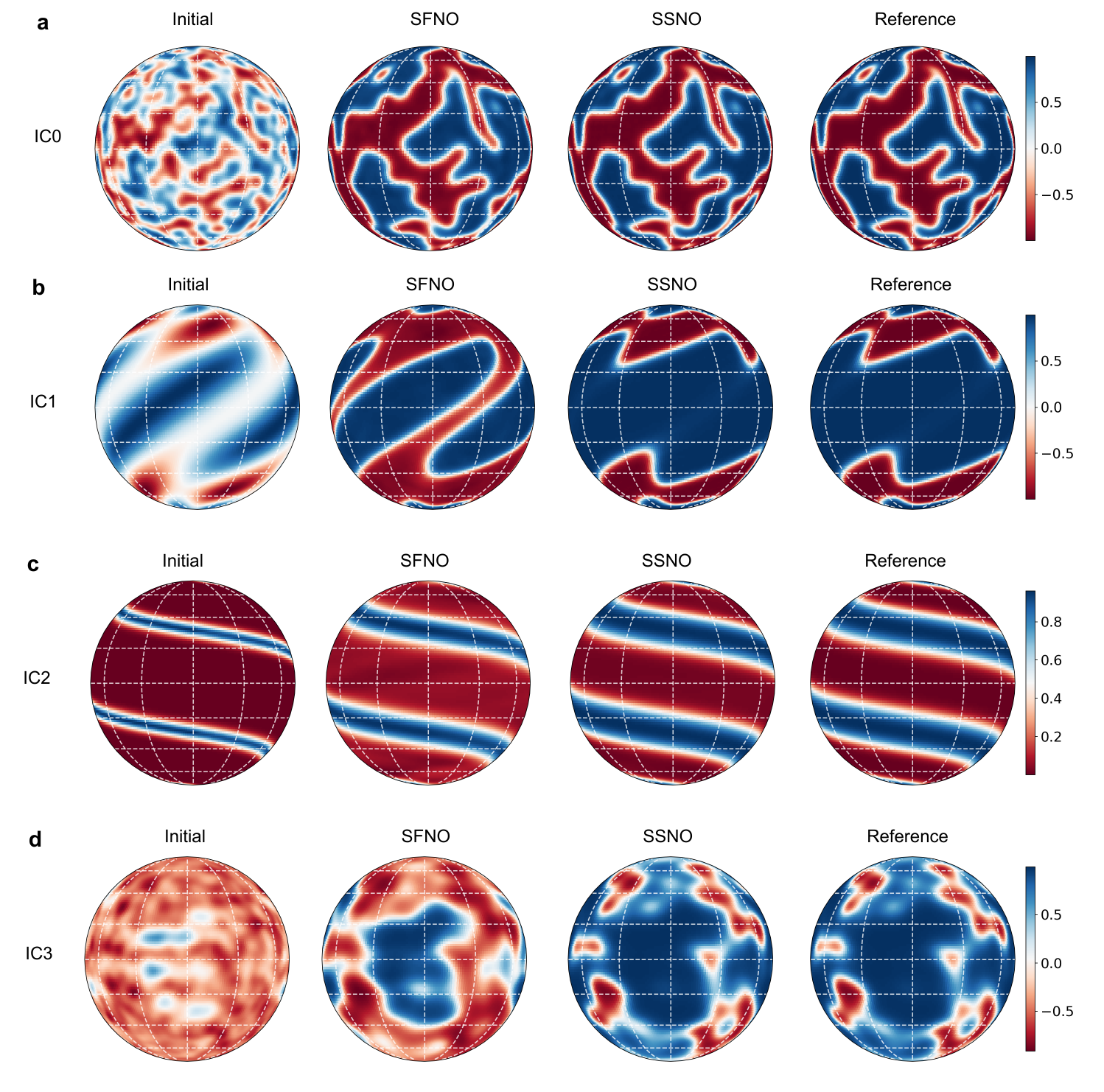}
    \caption{\textbf{Allen--Cahn on the sphere.} Snapshots at $t=5$ on the unit sphere $\mathbb{S}^2$ under four initial-condition families: \textbf{a}, IC0 (Gaussian random field; in-distribution). \textbf{b}, IC1 (band-limited zonal stripes; OOD). \textbf{c}, IC2 (oblique stripes; OOD). \textbf{d}, IC3 (localized spots; OOD). Columns show the initial field, SFNO, SSNO, and the reference solution. The predicted evolutions are further depicted in \textcolor{blue}{Supplementary Fig. S.6}.}
    \label{fig:4}  
\end{figure}

To probe non-Euclidean geometry and curvature effects, we study phase separation on the unit sphere governed by
\begin{equation}
\label{eq:ac_sphere}
\frac{\partial u}{\partial t} = \epsilon\,\Delta_{\mathbb{S}^2} u - \big(u^3 - u\big),
\end{equation}
where $\Delta_{\mathbb{S}^2}$ is the Laplace--Beltrami operator. We fix $\epsilon=10^{-3}$, which determines the diffusion strength and the effective interface thickness~\cite{allen1975coherent}. Relative to flat domains, curvature on $\mathbb{S}^2$ alters coarsening pathways and equilibrium patterns. The field is discretized on a $128\times256$ latitude-longitude grid; rollouts extend to $t=5$ with a fixed time step $\delta t=0.5$. Training uses five trajectories with Gaussian-random initial conditions (IC0). To test OOD generalization, we evaluate on three unseen families, including band-limited zonal stripes (IC1), oblique stripes (IC2), and localized spot fields (IC3).

Adapting SSNO to the sphere only requires a change of basis: we replace the FFT with a spherical-harmonic transform (SHT). The Freq2Vec module embeds spherical frequencies via harmonic degree and order $(\ell,m)$, and parameterizes derivative multipliers as functions of $(\ell,m)$. We adopt Spherical FNO (SFNO)~\cite{bonev2023spherical} as a baseline, which generalizes FNO to spherical geometries by replacing the neural Fourier layers with spherical-harmonic convolutions. 

Fig.~\ref{fig:4} compares predictions at $t=5$ across the four initial-condition families. For the in-distribution case (IC0), both SSNO and SFNO closely match the reference. In the OOD families, SFNO exhibits degradation: on zonal stripes (IC1), it exhibits phase drift and band bending; on oblique stripes (IC2), the interfaces broaden and lose sharpness, with misoriented bands and curvature errors; and on localized spots (IC3), small domains are over-smoothed or vanish, with spurious oscillations near interfaces. These visual failures are confirmed by quantitative metrics in Extended Data Fig.~\ref{fig:extendfig1}, where SFNO's relative $\ell_2$ error exceeds 1.0 and its correlation with the reference drops to approximately 0.74 on the most challenging OOD case. In contrast, SSNO preserves band orientation and phase, maintains interface width and curvature, and retains the size and placement of spots across IC1-IC3. Quantitatively, SSNO's superior performance is evidenced by a correlation coefficient consistently above 0.99 for all cases (Extended Data Fig.~\ref{fig:extendfig1}). We attribute this superior robustness to SSNO's spectrally inspired structures, which encourage SSNO to learn the underlying differential operators and achieve strong inductive biases rather than memorizing superficial patterns.

\subsection*{3D reaction--diffusion systems}

\begin{figure}[!t]
\centering
\includegraphics[width=1\linewidth]{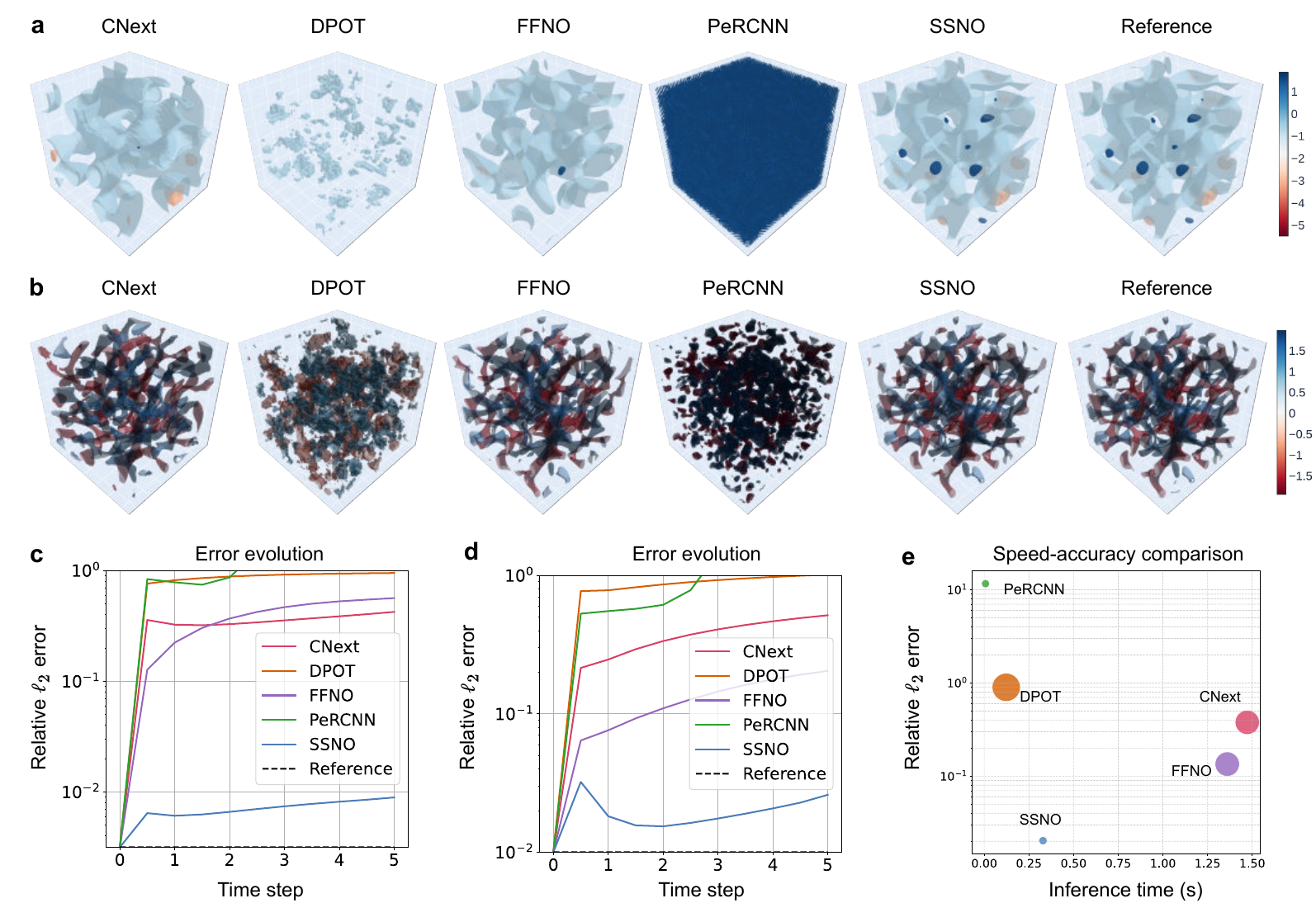}
\caption{\textbf{Multi-step error and predicted snapshots from SSNO and baselines on 3D reaction--diffusion systems.} \textbf{a} and \textbf{b}, Predicted fields for KSE and SHE at $t=5$. \textbf{c} and \textbf{d}, Multi-step relative $\ell_2$ error for KSE and SHE. \textbf{e}, Speed-accuracy comparison for SHE, where marker size represents the number of parameters of each model. The predicted evolutions for KSE and SHE are further depicted in \textcolor{blue}{Supplementary Figs. S.7--S.8}.}
\label{fig:5}
\end{figure}

We extend our study to three-dimensional reaction--diffusion dynamics, considering both the 3D KSE and the 3D SHE. The domain is $[0,8\pi)^3$ discretized on a uniform $64^3$ grid. We train on two trajectories and roll out to $T=5$ with a fixed time step $\delta t=0.5$. The baselines follow the 2D setting: DPOT, FFNO, CNext, and the physics-encoded PeRCNN.

Volume renderings at $t=5$ (Fig.~\ref{fig:5}\textbf{a}--\textbf{b}) show that SSNO preserves characteristic morphology in both systems, maintaining sharp interfaces and thin filaments while retaining large-scale organization. Among the baselines, PeRCNN exhibits large early errors at this time step and frequently diverges at later times. DPOT attains the lowest fidelity among the data-driven methods under this setting, with low correlations by $t=5$. FFNO and CNext reproduce large-scale patterns but distort fine-scale features; for example, both mislocate extrema and underestimate peak amplitudes for KSE, and they fail to capture thin connecting structures in SHE.

The relative $\ell_2$ errors over time (Fig.~\ref{fig:5}\textbf{c}--\textbf{d}) corroborate these observations. At the same time step $\delta t=0.5$, SSNO achieves the lowest error and remains stable over long horizons, improving upon the best baseline by about an order of magnitude on average (Fig.~\ref{fig:5}\textbf{c}--\textbf{d}). The 3D speed-accuracy comparison (Fig.~\ref{fig:5}\textbf{e}) places SSNO on the empirical Pareto frontier: at comparable runtime, it attains substantially lower error than FFNO, CNext, and PeRCNN. Overall, these results indicate that SSNO’s architectural principles scale to 3D stiff reaction--diffusion dynamics and remain data-efficient even with only two training trajectories.

\subsection*{Ablation studies}

\begin{figure}[!t]
    \centering
    \includegraphics[width=1.0\linewidth]{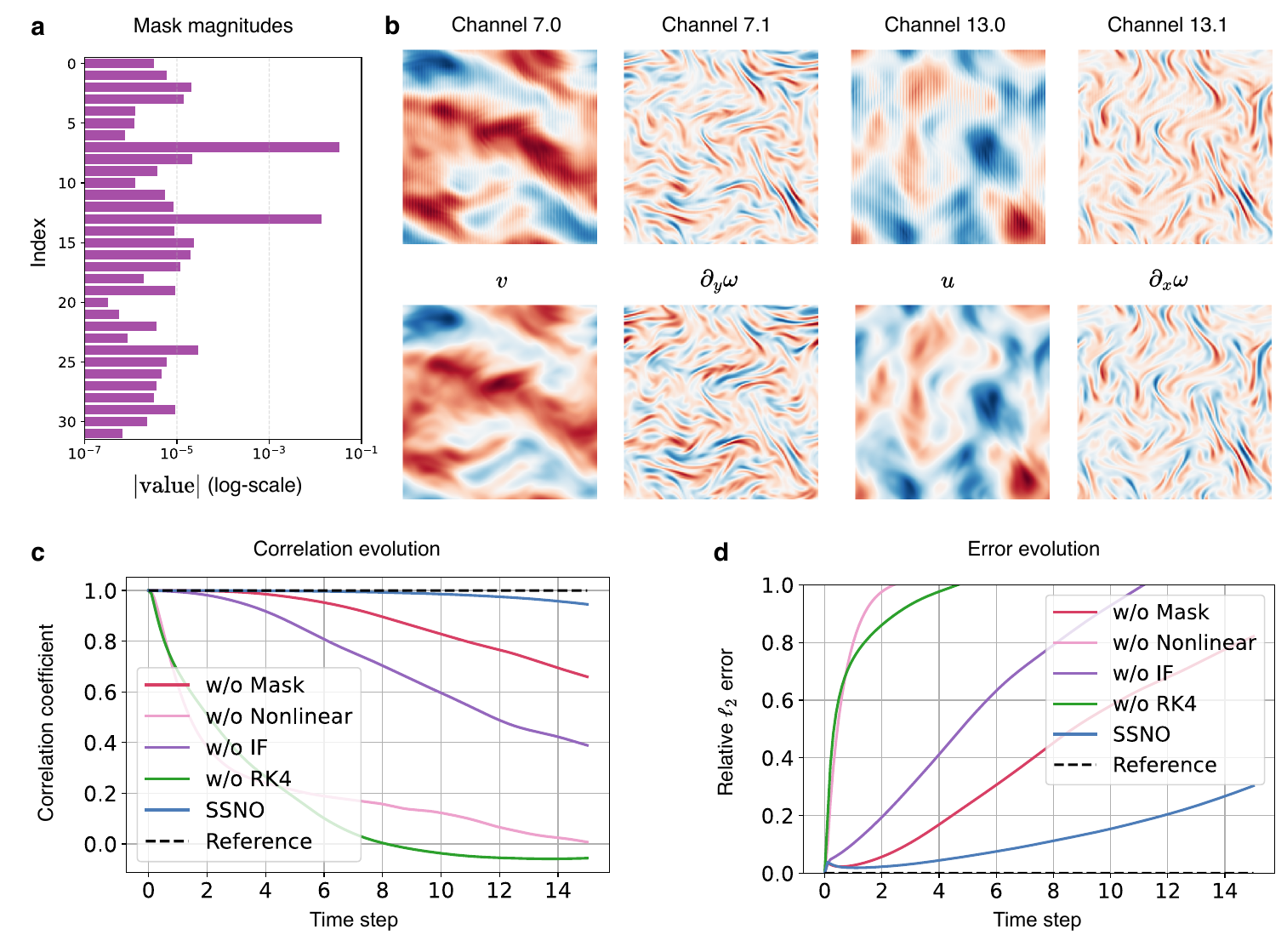}
  \caption{\textbf{Ablation studies and physical interpretability of SSNO on the 2D NSE ($\mathrm{Re}=1000$).} \textbf{a}, Learned sparse mask magnitudes in the $\Pi$ block. The mask is highly sparse, with two dominant indices (7 and 13) corresponding to the two multiplicative terms in the NSE. \textbf{b}, Comparison of feature fields from the two dominant channels. The top row displays the learned features from SSNO, which show a strong visual correspondence to the ground truth velocity components ($u, v$) and vorticity gradients ($\partial_x\omega, \partial_y\omega$) in the bottom row. \textbf{c} and \textbf{d}, Quantitative ablation results showing the evolution of the correlation coefficient (\textbf{c}) and relative $\ell_2$ error (\textbf{d}) for the full SSNO model versus ablated versions.}
\label{fig:6}
\end{figure}

We perform a series of ablation studies to dissect the SSNO architecture and quantify the contribution of its key components to the overall performance. These experiments are conducted on the 2D incompressible NSE benchmark with $\mathrm{Re}=1000$.

First, we investigate the physical interpretability of the learned operators, focusing on the sparse $\Pi$ block designed to model multiplicative nonlinearities. In the vorticity formulation of the 2D NSE, the advection term $\mathbf{u}\cdot\boldsymbol{\nabla} \omega$ contains two key multiplicative interactions, namely $u\partial_x\omega$ and $v\partial_y\omega$. We examined the learned sparse mask in the $\Pi$ block to see if it could recover this structure from data alone. As shown in Fig.~\ref{fig:6}\textbf{a}, the learned mask is indeed highly sparse, with the magnitudes of two channel indices (7 and 13) being several orders larger than the rest. To test whether these dominant channels correspond to the physical terms, we visualize their associated feature fields in Fig.~\ref{fig:6}\textbf{b}. The result reveals a remarkable correspondence: the learned features align directly with the velocity components ($u, v$) and the vorticity gradients ($\partial_x\omega, \partial_y\omega$), the constituent terms of the advection operator. This provides strong evidence that the sparsity prior guides SSNO to learn a physically meaningful and interpretable structure that mirrors the underlying unknown PDE.

For ablation studies, we remove one component at a time and evaluate long-term accuracy (Fig.~\ref{fig:6}\textbf{c}--\textbf{d}), using the same step size and training budget:
(i) \texttt{w/o IF}: replace the integrating factor treatment of the linear operator with an explicit update while retaining the same fourth-order staging;
(ii) \texttt{w/o RK4}: keep the integrating factor but replace the fourth-order IF-Runge--Kutta scheme with a first-order IF-Euler step;
(iii) \texttt{w/o Nonlinear}: remove the nonlinear blocks so only the learned linear operator remains; and
(iv) \texttt{w/o Mask}: drop the $\ell_1$ sparsity mask, allowing full channel mixing.
The full SSNO maintains high correlation and low relative $\ell_2$ throughout the rollout. Removing either the integrating factor or the RK4 staging destabilizes the time integration and leads to early divergence under the same $\delta t$, confirming their role in stabilizing stiff dynamics. Removing the nonlinear block causes a rapid loss of accuracy, consistent with the inability to represent advection. Eliminating the sparsity mask yields a gradual but consistent degradation, suggesting that the mask regularizes the model and helps identify the correct interaction structure from limited data.

Collectively, these studies demonstrate that all components of the SSNO architecture--the stable time-stepping scheme (IFRK4), the specialized nonlinear blocks, and the physics-inspired sparsity prior--are indispensable for achieving accurate, stable, and robust long-term predictions.

\section*{Discussion}

This paper introduces an equation-free learning framework (namely, SSNO) designed to model the dynamics of stiff PDE systems based on limited and sparsely sampled data. SSNO embeds spectrally inspired structures into its design to achieve strong inductive biases. Spatially, it operates in the frequency domain, using spectral transforms to capture global couplings and a novel Freq2Vec module to learn derivative-like operators directly from data without explicit PDE terms. This approach bypasses the locality constraints of convolutional FD and FV-based physics-encoded methods. Temporally, SSNO employs an IFRK4 scheme that analytically treats the stiff linear components of the dynamics. This design ensures numerical stability even with large time steps, a critical challenge in stiff systems. Furthermore, a sparsity prior is imposed on the learned nonlinear operators to promote a compact, PDE-like structure, enhancing data efficiency.

SSNO achieves state-of-the-art performance across a diverse suite of challenging 2D and 3D benchmarks in both Cartesian and spherical geometries. It consistently outperforms leading data-driven and physics-encoded baselines, yielding one to two orders of magnitude lower error in long-term predictions. A key advantage of SSNO is its remarkable data efficiency and generalization capability; it learns effectively from as few as two to five training trajectories and exhibits robust performance on OOD initial conditions where other models fail. Furthermore, our ablation studies reveal that SSNO is not a black box; the sparsity regularizer guides the model to learn an interpretable structure that directly corresponds to the physical operators of the underlying system. These results establish SSNO as a practical, data-efficient, and accurate neural solver for modeling and simulation in complex systems governed by unknown equations and observed via sparse data.

The present study suggests several avenues for our future work. On the theory side, it would be valuable to characterize the stability region and global error of the learned linear operator combined with IFRK4 time-stepping, and to analyze identifiability and generalization under few trajectories and sparse-in-time supervision. On the numerical side, SSNO could benefit from adaptive step sizing with embedded error estimators, other integrating factor variants for more severe stiffness, and learnable anti-aliasing filters that balance fidelity and stability. Finally, incorporating hard or soft physical constraints (incompressibility, mass/energy conservation) and addressing multi-physics systems (e.g., compressible flows and reactive transport) would broaden SSNO's applicability for more challenging tasks.

\section*{Methods}
We herein introduce the method of SSNO, the evaluation metrics, and the baseline models.

\subsection*{Problem setting and preliminaries}

\paragraph{Problem settings.}

We consider PDE-governed dynamics on a domain \(\Omega\subset\mathbb{R}^d\) over a time horizon \([0,T]\), of the general form in Eq.~\ref{eq:pde}, decomposed into a linear operator \(\mathcal{L}\), multiplicative interactions \(\Pi\), and local reaction terms \(\mathcal{R}\). We adopt a realistic yet challenging regime in which the governing PDE is unknown, and observations are both scarce (e.g., \(N\approx 2\text{--}5\) training trajectories) and sparsely sampled in time, with a large sampling interval \(\delta t\) between snapshots. Given the limited training set \(\mathcal{D}_{\mathrm{train}}=\{(\mathbf{u}_n(\cdot,t_m))_{m=0}^{M}\}_{n=1}^{N}\) with \(t_{m+1}-t_m=\delta t\), we aim to learn an autoregressive neural PDE surrogate \(\mathcal{G}_\theta(\cdot;\delta t)\), which advances the field across observation times:
\begin{equation}
    \mathbf{u}(\cdot,t_{m+1}) \approx \mathcal{G}_\theta\!\big(\mathbf{u}(\cdot,t_m);\delta t\big),~\label{eq:solver}
\end{equation}
enabling reliable long-horizon rollouts from limited data without requiring explicit PDE terms.

\paragraph{Preliminaries: spectral representation.}
Spectral methods employ a transform pair \((\mathcal{T},\mathcal{T}^{-1})\) to shuttle between physical and frequency domains. On periodic Cartesian grids, \(\mathcal{T}\) is the fast Fourier transform (FFT) with frequency index \(\mathbf{k}=(k_x,k_y,k_z)\); on the sphere, \(\mathcal{T}\) is a spherical-harmonic transform (SHT) indexed by degree/order \((\ell,m)\). In the frequency domain, spatial derivatives admit diagonal representations via simple multipliers, e.g.,
\begin{equation}
    \widehat{\boldsymbol{\nabla} _{x} \mathbf{u}}(\mathbf{k})= \mathrm{i} k_x \widehat{\mathbf{u}}(\mathbf{k}),\qquad
\widehat{\Delta \mathbf{u}}(\mathbf{k})= -\|\mathbf{k}\|^2_2 \widehat{\mathbf{u}}(\mathbf{k}),\qquad
\widehat{\Delta^2 \mathbf{u}}(\mathbf{k})= \|\mathbf{k}\|^4_2 \widehat{\mathbf{u}}(\mathbf{k}).\label{eq:derivative}
\end{equation}

Furthermore, some globally coupled operators admit concise representations in the frequency domain. For instance, in the vorticity formulation of 2D NSE (Eq.~\ref{eq:nse_vort}), recovering the velocity \(\mathbf{u}\) from vorticity \(\omega\) is a Poisson solve in physical space but reduces to per-mode algebra in Fourier space. With \(\mathbf{u}=\boldsymbol{\nabla} ^\perp\psi=(-\partial_y\psi,\partial_x\psi)\) and \(\omega=-\Delta\psi\), we have
\begin{equation}
\widehat{\psi}(\mathbf{k}) = -\frac{\widehat{\omega}(\mathbf{k})}{\|\mathbf{k}\|^2}, \quad
\widehat{\mathbf{u}}(\mathbf{k}) = \mathrm{i} \frac{\mathbf{k}^\perp}{\|\mathbf{k}\|^2} \widehat{\omega}(\mathbf{k}), \quad \text{with } \mathbf{k}^\perp := (-k_y, k_x).
\end{equation}

Compared with FD- or FV-based discretizations, spectral methods offer (i) {spectral accuracy}, for smooth solutions, errors decay nearly exponentially with resolution; (ii) a {global} representation that can handle global interactions; and (iii) {cheap, exact} differentiation/inversion via diagonal multipliers, realized with \(\mathcal{O}(N\log N)\) FFT/SHT calls~\cite{shen2011spectral, JML1268}.

\paragraph{Preliminaries: system stiffness.}

Stiffness classically refers to ordinary differential equation (ODE) systems whose components evolve on widely separated timescales, so that the fastest modes impose step-size constraints that are unnecessarily severe for the slow modes~\cite{higham1993stiffness}. In numerical PDEs, spatial discretization yields a system of ODEs, and often results in stiffness~\cite{krogstad2005generalized}.

This issue becomes particularly acute because spectral discretization transforms the PDE into a large system of coupled ODEs for the modal coefficients:
\begin{equation}
\frac{d}{dt}\,\widehat{\mathbf{u}}_{\mathbf{k}}(t)
= \lambda_{\mathbf{k}}\,\widehat{\mathbf{u}}_{\mathbf{k}}(t)
+ \widehat{\mathcal{N}}_{\mathbf{k}}\big({\mathbf{u}}(t)\big),    \label{eq:spectral_ode}
\end{equation}
where \(\mathcal{N}:=\Pi+\mathcal{R}\) collects the nonlinear interactions, and \(\lambda_{\mathbf{k}}\le 0\) is the symbol of the linear operator \(\mathcal{L}\) (scaling like \(-\|\mathbf{k}\|_2^2\) for diffusion or \(-\|\mathbf{k}\|_2^4\) for bi-Laplacian). The stiffness ratio
\begin{equation}
    \kappa_{\mathrm{stiff}}
:= \frac{\max_{\mathbf{k}}|\lambda_{\mathbf{k}}|}{\min_{\mathbf{k}\neq 0}|\lambda_{\mathbf{k}}|}
\end{equation}
reflects the severity of stiffness, which can be very large when high-frequency modes decay much faster than low modes. On a periodic box with \(N\) points per dimension, the stiffness ratio scales as \(\mathcal{O}(N^2)\) for diffusion and \(\mathcal{O}(N^4)\) for bi-Laplacian, respectively.

To accommodate stiffness induced by \(\mathcal{L}\), an integrating factor (exponential) formulation is employed~\cite{krogstad2005generalized}. Let \(\mathbf{w}_{\mathbf{k}}(t):=e^{-\lambda_{\mathbf{k}} t}\,\widehat{\mathbf{u}}_{\mathbf{k}}(t)\), we can transform Eq.~\ref{eq:spectral_ode} as follows
\begin{equation}
\frac{d}{dt}\mathbf{w}_{\mathbf{k}}(t)=e^{-\lambda_{\mathbf{k}} t}\,\widehat{\mathcal{N}}_{\mathbf{k}}\big({\mathbf{u}}(t)\big),
\end{equation}
where the linear term \(\lambda_{\mathbf{k}}\widehat{\mathbf{u}}_{\mathbf{k}}\) can be handled exactly. Integrating over one step \([t,\,t+\delta t]\) gives the following exact representation:
\begin{equation}
    \widehat{\mathbf{u}}_{\mathbf{k}}(t+\delta t)
= e^{\lambda_{\mathbf{k}}\delta t}\,\widehat{\mathbf{u}}_{\mathbf{k}}(t)
+ e^{\lambda_{\mathbf{k}}(t+\delta t)}\!\!\int_{t}^{t+\delta t}\!
e^{-\lambda_{\mathbf{k}}\tau}\,\widehat{\mathcal{N}}_{\mathbf{k}}\big(\mathbf{u}(\tau)\big)\,d\tau.\label{eq:int}
\end{equation}
An integrating factor fourth-order Runge--Kutta scheme (IFRK4) approximates Eq.~\ref{eq:int} with fourth-order accuracy while preserving the exact exponential propagation of the linear part~\cite{krogstad2005generalized}. As a result, the dominant linear CFL restriction is removed: stability is controlled primarily by the nonlinear term \(\mathcal{N}\), allowing much larger time steps than fully explicit methods. 

\subsection*{Architecture of SSNO}

\paragraph{Overall architecture.}
We propose SSNO to learn stiff dynamics from scarce, sparsely sampled data without explicit PDE terms. A single step advances \(\mathbf{u}_m \!\to\! \mathbf{u}_{m+1}\) through the following three components (Fig.~\ref{fig:1}\textbf{a}).

First, SSNO maps the state to the frequency domain via a transform pair \((\mathcal{T},\mathcal{T}^{-1})\): FFTs on Cartesian grids and an SHT on the sphere (Fig.~\ref{fig:1}\textbf{b}). A lightweight Frequency-to-Vector module (Freq2Vec) embeds the frequency index \(\mathbf{k}\) and outputs per-mode multipliers. This yields a learned linear symbol \(\lambda_{\phi}(\mathbf{k})\) and derivative-like spectral features (Fig.~\ref{fig:1}\textbf{c}).

Second, the nonlinear effects are decomposed into a local reaction block (Fig.~\ref{fig:1}\textbf{d}) and a nonlocal \(\Pi\) block (Fig.~\ref{fig:1}\textbf{e}). The reaction block \(\mathcal{R}_{\eta}\) is a point-wise MLP in physical space. The \(\Pi\) block \(\Pi_{\zeta}\) models nonlinear interactions (e.g., convection operator $\mathbf{u} \cdot \boldsymbol{\nabla}{\mathbf{u}}$) via multiplication. SSNO computes only a sparse set of products selected by a learnable mask with \(\ell_1\) regularization, applies a low-pass filter to suppress aliasing, thus realizing a compact yet expressive interaction pattern.

Third, time marching adopts an IFRK4 scheme (Fig.~\ref{fig:1}\textbf{f}): the stiff linear part \(\mathcal{L}_{\phi}\) is propagated analytically via exponential computation, while the nonlinear term is evaluated at explicit stages.

In summary, SSNO embeds spectrally inspired structures into the architecture, providing strong inductive biases that enable learning from few trajectories and robust OOD generalization. Spatially, operating in the frequency domain allows the model to learn both local and global interactions and to compute high-fidelity spatial derivatives, outperforming FD/FV-based alternatives. Temporally, an integrating factor scheme propagates the stiff linear dynamics analytically while treating nonlinearities explicitly, yielding stable long-term rollouts with large \(\delta t\).

\paragraph{Linear block \(\mathcal{L}_{\phi}\).}
The linear block \(\mathcal{L}_{\phi}\) works in the frequency domain with \(\widehat{\mathbf{u}}=\mathcal{T}[\mathbf{u}]\), where \(\mathbf{u}\) can be the physical field or an intermediate IFRK4 state. 

To learn per-mode multipliers of spatial derivatives (e.g., $-\|\mathbf{k}\|_2^2$ in Eq.~\ref{eq:derivative}), the {Freq2Vec}\(_L\) module implemented as a small MLP maps \(\mathbf{k}\) to two branches \({y}_\mathbf{k}\) and \({z}_\mathbf{k}\), defining the complex symbol
\begin{equation}
\lambda_{\phi}(\mathbf{k}) \;=\; -\,\mathrm{softplus}({y}_\mathbf{k}) \;+\; \mathrm{i}\, {z}_\mathbf{k},
\label{eq:ssno_lin_symbol}
\end{equation}
\begin{equation}
\text{with}\quad\mathrm{softplus}({x}) \;=\; \log\!\big(1+e^{{x}}\big) \;>\; {0}\quad \forall\,{x}.
\label{eq:ssno_softplus}
\end{equation}

The positivity of $\mathrm{softplus}(\cdot)$~\cite{zheng2015improving} enforces  \(\operatorname{Real}[\lambda_{\phi}(\mathbf{k})]\le 0\), encoding a dissipative prior and keeping the integrating factors \({E}^{\delta t}(\mathbf{k})=\exp\!\big(\lambda_{\phi}(\mathbf{k})\,\delta t\big)\) bounded:
\begin{equation}
\big|E^{\delta t}(\mathbf{k})\big|
= \Big|\exp\!\big(\lambda_{\phi}(\mathbf{k})\,\delta t\big)\Big|
= \exp\!\big(\delta t\,\operatorname{Real}[\lambda_{\phi}(\mathbf{k})]\big)
\le 1,\qquad \forall\ \delta t\ge 0,
\label{eq:if_bound}
\end{equation}
thereby improving numerical stability during both training and inference.

The linear update is then realized by per-mode multiplication followed by an inverse transform:
\begin{equation}
\widehat{\mathcal{L}_{\phi}[\mathbf{u}]}(\mathbf{k};\delta t)
\;=\; E^{\delta t}(\mathbf{k})\,\widehat{\mathbf{u}}(\mathbf{k}),
\qquad
\mathcal{L}_{\phi}(\mathbf{u};\delta t)
\;=\; \mathcal{T}^{-1}\!\big[\widehat{\mathcal{L}_{\phi}\mathbf{u}}\big].
\end{equation}

Overall, the linear block $\mathcal{L}_{\phi}$ advances the stiff linear dynamics analytically, with computational cost dominated by FFT/SHT calls of order \(\mathcal{O}(N\log N)\).

\paragraph{Reaction block \(\mathcal{R}_{\eta}\).}

Reaction terms are ubiquitous in physical PDEs and typically appear as local nonlinear maps \(f(\mathbf{u})\) applied point-wise, such as the Allen--Cahn nonlinearity. These terms act locally at each point without coupling different spatial locations, capturing saturation, threshold effects, and other site-specific kinetics~\cite{halatek2018rethinking}.

To model these local effects, we use a shallow point-wise MLP in physical space:
\begin{equation}
    \mathcal{R}_{\eta}[\mathbf{u};\delta t]=\mathrm{MLP}_{\eta} \big(\mathbf{u}\big)\delta t.
\end{equation}
with weights shared across spatial locations. This captures common polynomial and saturating behaviors at negligible cost, is agnostic to grid or geometry, and does not interfere with the global couplings handled by the spectral blocks.

\paragraph{$\Pi$ block \(\Pi_{\zeta}\).}
Many PDE nonlinearities are products between the state and its spatial derivatives (e.g., the convection terms or velocity-vorticity couplings). To model these interactions without hand-crafted terms, \(\Pi_{\zeta}\) builds two derivative-like feature streams in the frequency domain, multiplies corresponding pairs in physical space to simulate nonlinearities, and regularizes the result with sparsity and anti-aliasing.

We first transform the physical field $\mathbf{u}$ to the frequency domain, i.e., \(\widehat{\mathbf{u}}=\mathcal{T}[\mathbf{u}]\). Similar to the linear block, a lightweight {Freq2Vec}\(_\Pi\) maps each frequency index \(\mathbf{k}\) to a bank of per-mode multipliers
\(\{\alpha_r(\mathbf{k})\}_{r=1}^{R}\).
These multipliers act per-mode to produce derivative-like spectral features
\begin{equation}
\widehat{\mathbf{z}}_r(\mathbf{k}) \;=\; \alpha_r(\mathbf{k})\,\widehat{\mathbf{u}}(\mathbf{k}),\qquad r=1,\dots,R.
\label{eq:z_bank}
\end{equation}
Two separate \(1{\times}1\) convolutional layers then form the two branches from this shared bank:
\begin{equation}
\widehat{\mathbf{v}}_i(\mathbf{k}) \;=\; \sum_{r=1}^{R} \mathbf{W}^{(v)}_{i,r}\,\widehat{\mathbf{z}}_r(\mathbf{k}),\quad
\widehat{\mathbf{w}}_i(\mathbf{k}) \;=\; \sum_{r=1}^{R} \mathbf{W}^{(w)}_{i,r}\,\widehat{\mathbf{z}}_r(\mathbf{k}),\quad i=1,\dots,Q,
\label{eq:two_branches}
\end{equation}
where \(\mathbf{W}^{(v)}\) and \(\mathbf{W}^{(w)}\) are the \(1{\times}1\) channel-mixing weight matrices.
After inverse transforms, we obtain
\(\mathbf{v}_i(\mathbf{x})=\mathcal{T}^{-1}[\widehat{\mathbf{v}}_i(\mathbf{k})]\) and
\(\mathbf{w}_j(\mathbf{x})=\mathcal{T}^{-1}[\widehat{\mathbf{w}}_j(\mathbf{k})]\).

To simulate nonlinearity, we multiply corresponding pairs and gate them with a learnable mask \(\mathbf{m}\in\mathbb{R}^{Q}\):
\begin{equation}
\mathbf{s}_i(\mathbf{x}) \;=\; \mathbf{m}_i\, \mathbf{v}_i(\mathbf{x})\, \mathbf{w}_i(\mathbf{x}), \qquad i=1,\ldots,Q.
\label{eq:masked_pair_pi}
\end{equation}
Because PDE nonlinearities are often algebraically compact (e.g., few multiplicative terms), we impose sparsity via $\ell_1$ regularization on the mask to reflect this prior. Despite its simplicity, this gated product mechanism captures a broad class of nonlinear couplings, making the \(\Pi\) block both compact and physically consistent.

However, such multiplicative interactions inject energy above the Nyquist limit, leading to aliasing that contaminates low frequencies~\cite{kravchenko1997effect}. We therefore de-alias each product channel in the spectral domain using the classical \(2/3\) low-pass rule:
\begin{equation}
\widehat{\mathbf{s}}^{\,\mathrm{filt}}_i(\mathbf{k})
=\Lambda(\mathbf{k})\,\mathcal{T}[\mathbf{s}_i](\mathbf{k}),\qquad
\Lambda(\mathbf{k})=
\begin{cases}
1,& \|\mathbf{k}\|_\infty \le \tfrac{2}{3}k_{\max},\\
0,& \text{otherwise}.
\end{cases}
\label{eq:deal_s_i}
\end{equation}
After the inverse transform \(\mathbf{s}_i^{\mathrm{filt}}=\mathcal{T}^{-1}[\widehat{\mathbf{s}}^{\,\mathrm{filt}}_i]\), we aggregate the \(Q\) channels with a \(1{\times}1\) convolution to obtain the \(\Pi\) block output:
\begin{equation}
\Pi_{\zeta}[\mathbf{u};\delta t]
\;=\;
\mathrm{Conv}_{1\times 1}\!\big([\mathbf{s}_1^{\mathrm{filt}},\ldots,\mathbf{s}_Q^{\mathrm{filt}}]\big)\;\delta t.
\label{eq:pi_agg}
\end{equation}
The de-aliasing filter prevents spurious spectral leakage, thereby improving numerical stability when representing nonlinear terms.

\paragraph{IFRK4 time-stepping.}
To handle the system stiffness, we use the classical IFRK4 scheme where the learned linear part is propagated analytically and the nonlinear part is evaluated explicitly.

Given \(\mathbf{u}_m\) at time \(t_m\), the stage increments in IFRK4 scheme~\cite{krogstad2005generalized} are calculated as follows:
\begin{equation}
\begin{aligned}
\mathbf{a}_m &= \mathcal{N}_\theta[\mathbf{u}_m, \delta t],\\
\mathbf{b}_m &= \mathcal{N}_\theta\big[\mathcal{L}_{\phi}[\mathbf{u}_m+\tfrac{1}{2}\mathbf{a}_m;\tfrac{\delta t}{2}]; \tfrac{\delta t}{2}]\big],\\
\mathbf{c}_m &= \mathcal{N}_\theta\big[\mathcal{L}_{\phi}[\mathbf{u}_m;\tfrac{\delta t}{2}]+\tfrac{\mathbf{b}_m}{2}; \tfrac{\delta t}{2}]\big],\\
\mathbf{d}_m &= \mathcal{N}_\theta\big[\mathcal{L}_{\phi}[\mathbf{u}_m;{\delta t}]+\mathcal{L}_{\phi}[\mathbf{c}_m;\tfrac{\delta t}{2}]; \tfrac{\delta t}{2}\big],
\end{aligned}
\label{eq:ifrk4_stages_correct}
\end{equation}
where \(\mathcal{N}_\theta \;:=\; \mathcal{R}_{\eta} \;+\; \Pi_{\zeta}\) denotes the nonlinear module. The next state is
\begin{equation}
\mathbf{u}_{m+1}
= \mathcal{L}_{\phi}[\mathbf{u}_m;\delta t]
+\frac{1}{6}\Big(
\mathcal{L}_{\phi}[\mathbf{a}_m;\delta t]
+2\mathcal{L}_{\phi}[\mathbf{b}_m+\mathbf{c}_m;  \tfrac{\delta t}{2}]
+\mathbf{d}_m\Big).
\label{eq:ifrk4_update_correct}
\end{equation}
Eqs. \ref{eq:ifrk4_stages_correct} and \ref{eq:ifrk4_update_correct} match the standard IFRK4 form with \(N\equiv\mathcal{N}_\theta\) and \(L\equiv\mathcal{L}_\phi\). This time-stepping scheme eliminates the dominant linear CFL restriction, allowing for stable long-term rollouts at large \(\delta t\).

\subsection*{Evaluation metrics}

We use two primary metrics to quantitatively evaluate the prediction accuracy of each model, including the relative $\ell_2$ error and correlation coefficient.

The relative \(\ell_2\) error measures the normalized difference between the predicted solution \({\mathbf{u}}_{\operatorname{pre}}\) and the reference solution \(\mathbf{u}\). It is defined as
\[
\text{Relative } \ell_2 \text{ error} = \frac{\| {\mathbf{u}_{\operatorname{pre}}} - \mathbf{u} \|_2}{\| \mathbf{u} \|_2} = \frac{\sqrt{\sum_i ({\mathbf{u}_{\operatorname{pre}}}_{(i)} - \mathbf{u}_{(i)})^2}}{\sqrt{\sum_i \mathbf{u}_{(i)}^2}},
\]
where the sums run over all spatial points in the solution.

The correlation coefficient evaluates the linear correlation between the predicted and reference solutions, defined by
\[
\text{Corr}({\mathbf{u}_{\operatorname{pre}}}, \mathbf{u}) = \frac{\sum_i ({\mathbf{u}_{\operatorname{pre}}}_{(i)} - \overline{{\mathbf{u}}}_{\operatorname{pre}})(\mathbf{u}_{(i)} - \bar{\mathbf{u}})}{\sqrt{\sum_i ({\mathbf{u}_{\operatorname{pre}}}_{(i)} - \overline{{\mathbf{u}}}_{\operatorname{pre}})^2} \sqrt{\sum_i (\mathbf{u}_{(i)} - \bar{\mathbf{u}})^2}},
\]
where \(\overline{{\mathbf{u}}}_{\operatorname{pre}}\) and \(\bar{\mathbf{u}}\) denote the mean values of \({\mathbf{u}}_{\operatorname{pre}}\) and \(\mathbf{u}\), respectively.

\subsection*{Baseline models}

For 2D cases, we benchmark SSNO against state-of-the-art neural PDE solvers, including the Fourier-based neural operators DPOT~\cite{hao2024dpot} and FFNO~\cite{tran2023factorized}, the convolutional-based PDE solver CNext~\cite{liu2022convnet,ohana2024well}, the transformer-based PDE solver FactFormer~\cite{li2023scalable}, and the physics-encoded model PeRCNN~\cite{rao2023encoding}, which embeds FD stencils into its architecture. For 3D cases, the set of baseline models is consistent with the 2D setting, including DPOT, FFNO, CNext, and the physics-encoded PeRCNN. For the ACE on the sphere, we adopt Spherical FNO (SFNO)~\cite{bonev2023spherical} as a baseline, which generalizes FNO to spherical geometries by replacing the Fourier transformation with spherical harmonic convolutions. 

\section*{Data availability} 
All the used datasets are available on GitHub at \url{https://github.com/intell-sci-comput/SSNO}.

\section*{Code availability} 
All the source codes to reproduce the results in this study are available on GitHub at \url{https://github.com/intell-sci-comput/SSNO}.

\bibliographystyle{unsrt}
\bibliography{references}

\vspace{36pt}
\noindent\textbf{Acknowledgement:}
The work is supported by the National Natural Science Foundation of China (No. 62276269, No. 62506367, and No. 92270118), the Beijing Natural Science Foundation (No. 1232009), and the Strategic Priority Research Program of the Chinese Academy of Sciences (No. XDB0620103). R.Z. would like to acknowledge the supported by the China Postdoctoral Science Foundation under Grant Number 2025M771582 and the Postdoctoral Fellowship Program of CPSF under Grant Number GZB20250408. In addition, Y.L. would like to acknowledge the support from the Fundamental Research Funds for the Central Universities (E2EG2202X2). \\

\noindent\textbf{Author contributions:} R.Z. and H.S. contributed to the ideation and design of the research; R.Z. conducted the experiments of SSNO and baseline models; H.S. supervised the project; all authors contributed to the research discussions, writing, and editing of the paper. \\

\noindent\textbf{Correspondence to:} Hao Sun (\url{haosun@ruc.edu.cn}).\\

\noindent\textbf{Competing interests:}
The authors declare no competing interests.\\

\noindent\textbf{Supplementary information:}
The supplementary information is attached.


\clearpage
\setcounter{figure}{0}
\renewcommand{\figurename}{Extended Data Figure}
\setcounter{table}{0}
\renewcommand{\tablename}{Extended Data Table}


\begin{table*}[!h]
\centering
\small
\caption{\textbf{Comparison of method families under practical data constraints for stiff PDE dynamics.}
We highlight (i) \textbf{Equation-free}: no access to PDE terms, (ii) \textbf{Few trajectories}: only a small number of long sequences are available (e.g., 2--5),
(iii) \textbf{Sparse-in-time}: can conduct simulation or learning with a large sampling interval $\delta t$, and (iv) \textbf{Global spatial coupling}: built-in long-range spatial interactions. `NA' indicates not applicable.}
\label{tab:extendtab1}
\resizebox{\textwidth}{!}{
\begin{tabular}{lcccc}
\toprule
\textbf{Method family} 
& \textbf{Equation-free} 
& \textbf{Few trajectories} 
& \textbf{Sparse-in-time} 
& \textbf{Global spatial coupling}\\
\midrule
Classical spectral methods~\cite{shen2011spectral}   & \xmark & NA   & \xmark   & \cmark \\
integrating factor schemes~\cite{curtiss1952integration} & \xmark & NA   & \cmark   & \cmark \\
Data-driven methods~\cite{li2021fourier}             & \cmark & \xmark& \xmark& \cmark \\
Physics-informed methods~\cite{li2024physics}        & \xmark & \cmark& \xmark& \cmark \\
FD/FV-based methods~\cite{rao2023encoding}           & \xmark & \cmark& \xmark& \xmark \\
Spectral-based methods~\cite{li2025symbolic}         & \xmark & \cmark& \xmark& \cmark \\
\midrule
\textbf{SSNO} (ours)                                  & \cmark & \cmark& \cmark& \cmark \\
\bottomrule
\end{tabular}}
\end{table*}

\clearpage
\begin{table*}[!h]
\centering
\caption{\textbf{Performance of SSNO on the 2D KSE across different rollout time step sizes ($\delta t$).} The table shows the average relative $\ell_2$ error and the correlation coefficient over time.}
\label{tab:extendtab2}
\begin{tabular}{@{}lcccc@{}}
\toprule
\textbf{Metric} & ${\delta t = 0.1}$ & ${\delta t = 0.2}$ & ${\delta t = 0.5}$ & ${\delta t = 1.0}$ \\ \midrule
Relative $\ell_2$ error & $1.10 \times 10^{-3}$ & $1.20 \times 10^{-3}$ & $2.15 \times 10^{-3}$ & $5.55 \times 10^{-3}$ \\
Correlation coefficient & 0.9999938 & 0.9999910 & 0.9999737 & 0.9998180 \\ \bottomrule
\end{tabular}
\end{table*}

\clearpage

\begin{figure}[!h]
  \centering
   \includegraphics[width=1\linewidth]{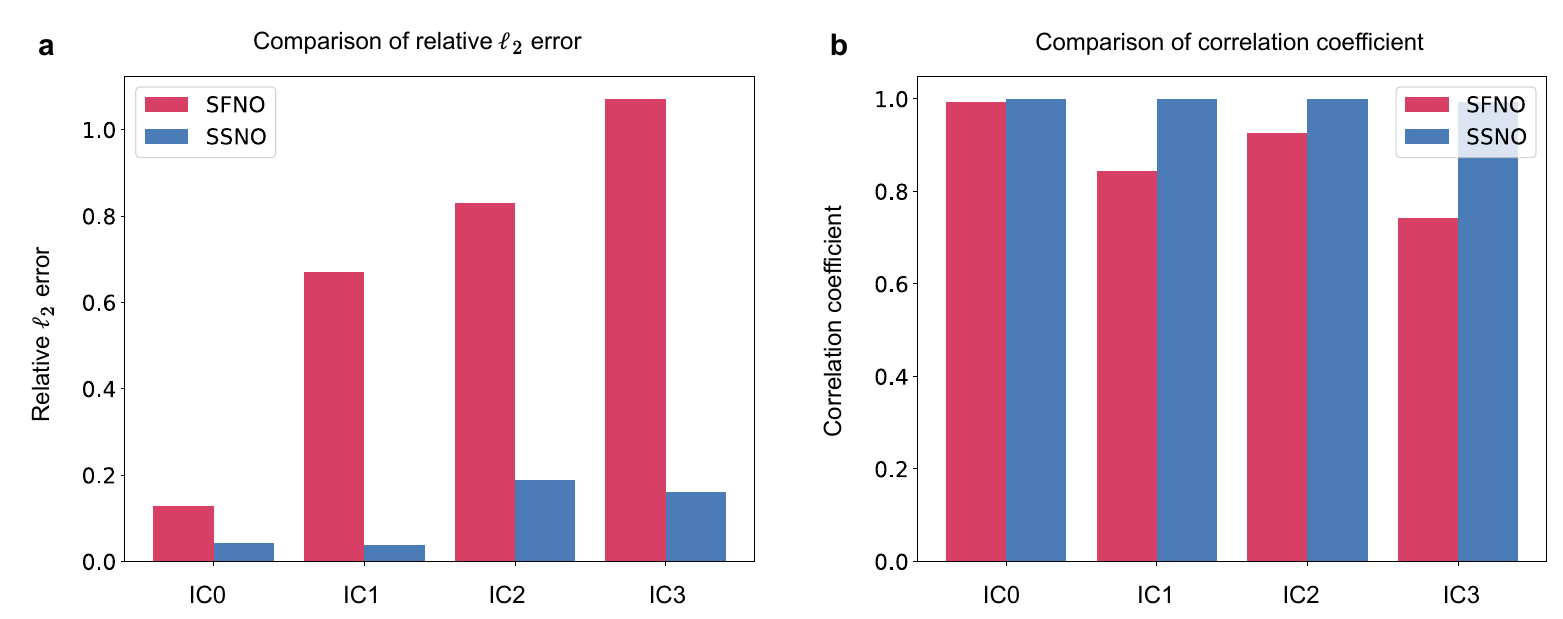}
\caption{\textbf{Quantitative comparison of SSNO and SFNO for the Allen-Cahn equation on the sphere.} Comparison of prediction accuracy at the final time $t=5$ across the four initial conditions (IC0: in-distribution; IC1--3: OOD): \textbf{a}, relative $\ell_2$ error (lower is better); \textbf{b}, correlation coefficient with the reference solution (higher is better).}
   \label{fig:extendfig1}
\end{figure}

\clearpage

\setcounter{figure}{0}
\renewcommand{\figurename}{Appendix Figure}
\setcounter{table}{0}
\renewcommand{\tablename}{Appendix Data Table}

\noindent This supplementary document provides a detailed description of the data generation process, details on implementing the Stable Spectral Neural Operator (SSNO) and baseline methods, and additional experimental results.

\section*{Supplementary Note A: Data generation and numerical setup}

Except for the spherical case, all problems are posed on periodic Euclidean domains and are discretized with Fourier spectral methods on grids of size \(N^{d}\) (with \(d\in\{2,3\}\)). Spatial derivatives are applied in the frequency domain, where nonlinear terms are evaluated in physical space and transformed back. We apply the standard \(2/3\) de-aliasing rule by zeroing high-wavenumber modes. Time series are stored at uniform intervals, including the initial state. Initial conditions on flat domains are sampled from an isotropic Gaussian random field. The spherical problem is posed on the unit sphere \(\mathbb{S}^{2}\); initial conditions are synthesized from random real spherical-harmonic coefficients. For each dataset, Table~\ref{tab:data_gen} reports the domain size, grid resolution, time step \(\delta t\), total simulated time \(T\), and the number of training/validation/test trajectories. Fixed random seeds are used for data generation: seed 0 for training, seed 1 for testing, and seed 2 for validation.

\begin{table}[t!]
\centering
\begin{threeparttable}
\caption{Computational parameters for dataset generation.}
\label{tab:data_gen}
\small
\begin{tabular}{lccccc}
\toprule
\textbf{Dataset} & Domain size & Grid resolution & $\delta t$ & $T$ & \#\,Train/Val./Test \\
\midrule
2D KSE & $[0,8\pi)^2$   & $128\times128$                 & $1\times10^{-5}$ & $5$        & 5/2/5 \\
3D KSE & $[0,8\pi)^3$  & $128\times128\times128$        & $4\times10^{-6}$ & $5$        &  2/2/5 \\
2D SHE & $[0,8\pi)^2$   & $128\times128$                 & $1\times10^{-5}$ & $5$        &  5/2/5 \\
3D SHE & $[0,8\pi)^3$  & $128\times128\times128$        & $1\times10^{-5}$ & $5$        &  2/2/5 \\
2D NSE & $[0,2\pi)^2$   & $512\times512$                 & $1\times10^{-4}$ & $8/15^{\dagger}$ &  5/2/5 \\
ACE on $\mathbb{S}^2$ & radius $1$ & $(N_\theta,N_\phi)=(256,512)$ & $1\times10^{-3}$ & $5$ &  5/2/5 \\
\bottomrule
\end{tabular}
\begin{tablenotes}[flushleft]
\footnotesize
\item $^{\dagger}$ Only the first $8$ time units are provided for training; testing evaluates extrapolation up to $15$.
\end{tablenotes}
\end{threeparttable}
\end{table}

\subsection*{2D/3D Kuramoto--Sivashinsky (KSE)}
The KSE, originally derived for laminar flame-front instabilities~\cite{sivashinsky1980flame} and widely used as a model of spatiotemporal chaos~\cite{michelson1977nonlinear}, takes the following nondimensional isotropic form:
\begin{equation}
\label{eq:kse}
\frac{\partial u}{\partial t} + \Delta u + \Delta^2 u + \frac{1}{2}\,\|\boldsymbol{\nabla}  u\|_2^2 = 0.
\end{equation}
The linear operator \(\Delta+\Delta^{2}\) is applied spectrally. The gradient \(\boldsymbol{\nabla}  u\) and the quadratic nonlinearity are computed in physical space and returned to Fourier space with \(2/3\) de-aliasing. Time integration employs the classical fourth-order Runge--Kutta (RK4) method. The initial condition is sampled from a zero-mean Gaussian random field.

\subsection*{2D/3D Swift--Hohenberg (SHE)}
The SHE models pattern-forming instabilities arising from finite-wavelength bifurcations~\cite{peletier2004pattern},
\begin{equation}
\label{eq:she}
\frac{\partial u}{\partial t} = u - (1+\Delta)^2 u - u^3,
\end{equation}
which selects a characteristic length scale and generates self-organized periodic structures. The linear term \(-(1+\Delta)^2 u\) and the cubic nonlinearity are handled via a standard splitting scheme. Time stepping uses RK4. The initial condition is sampled from a zero-mean Gaussian random field.

\subsection*{2D incompressible Navier--Stokes (NSE)}
To probe global coupling and long-term extrapolation, we consider the 2D vorticity equation
\begin{equation}
\begin{aligned}
    \frac{\partial \omega}{\partial t} + \mathbf{u}\cdot\boldsymbol{\nabla}  \omega 
    &= \frac{1}{\mathrm{Re}}\Delta \omega + f, \\
\boldsymbol{\nabla} \cdot\mathbf{u} &= 0, \\
\omega &= \boldsymbol{\nabla} \times\mathbf{u}, \label{eq:nse}
\end{aligned}
\end{equation}
with a stationary Kolmogorov-type vorticity forcing (the curl of a sinusoidal body force), e.g. \(f(\mathbf{x})=0.1\sin(8x_1)\). We vary the Reynolds number \(\mathrm{Re}\in\{500,1000,1500\}\) by adjusting the viscosity \(\nu\) to span laminar to turbulent regimes. At each step, we recover the velocity from vorticity through a streamfunction: solve \(\Delta\psi=\omega\) in Fourier space, then set \(\mathbf{u}=\boldsymbol{\nabla} ^\perp\psi\). Advection \(\mathbf{u}\cdot\boldsymbol{\nabla} \omega\) is formed in physical space with \(2/3\) de-aliasing. We use a Crank--Nicolson scheme for time-advancing, in line with the data generation regime in FNO~\cite{li2021fourier}. The initial condition is sampled from a zero-mean Gaussian random field.

\subsection*{Allen--Cahn on the sphere (ACE)}
To probe non-Euclidean geometry and curvature effects, we study phase separation on the unit sphere governed by
\begin{equation}
    \frac{\partial u}{\partial t} = \epsilon\,\Delta_{\mathbb{S}^2} u - \big(u^3 - u\big),\label{eq:ac_sphere}
\end{equation}
with \(\epsilon=10^{-3}\). The Laplace--Beltrami operator is diagonal in the real spherical-harmonic basis with eigenvalues \(-\ell(\ell+1)\) on the unit sphere. We use differentiable spherical-harmonic transforms (\texttt{torch-harmonics}~\cite{bonev2023spherical}) to shuttle between coefficient and spatial grids. Time stepping uses a two-step Adams--Bashforth scheme. To reduce sensitivity to high-frequency content at \(t=0\), we include a short warm-up run without recording and then enter the main recording window. Initial conditions are constructed by drawing spherical harmonic coefficients with a prescribed amplitude profile and transforming them back to the grid.

\section*{Supplementary Note B: Baseline methods and implementation details}

We herein introduce the baseline methods and implementation details.

\subsection*{Training details}

All models are trained and evaluated on NVIDIA A100 GPUs with PyTorch~\cite{NEURIPS2019_9015} as the deep learning framework. We fix the random seed to 0 for reproducibility. During training, we use Adam~\cite{kingma2015adam} with weight decay and a OneCycle learning-rate scheduler. At validation checkpoints, we save the model with the lowest validation relative error and report final metrics on the test data using this checkpoint. 

\paragraph{Teacher forcing and rollout.}
Let $n_1$ be the maximum number of warm-up (teacher-forced) steps and $n_2$ the supervised rollout horizon. At each iteration, we sample a mini-batch of with $B$ trajectories and uniformly random start times. We draw an integer $n\in\{0,\dots,n_1\}$; if $n>0$, the network first performs an $n$-step closed-loop warm-up from the single observed frame to produce the input state at $t{+}n$. We then unroll the model for $n_2$ steps in a closed loop and supervise these predictions against the ground-truth frames that follow the start time by an offset of $n{+}1$. This schedule mixes short- and medium-horizon supervision and stabilizes long rollouts~\cite{brandstetter2022message}.

\paragraph{Loss and optional sparsity.}
The training objective is an $\ell_2$ root-mean-square error over the predicted rollout:
\[
Loss \;=\; \sqrt{\frac{1}{B\,n_2} \sum_{b,t}\!\| \hat{\mathbf{u}}^{(b)}_{t} - \mathbf{u}^{(b)}_{t} \|_{2}^{2} + \varepsilon}\; +\; \beta\,\|\mathbf{m}\|_{1},
\]
where $\varepsilon=10^{-7}$ ensures numerical stability and the $\ell_1$ term is included only when a sparsity coefficient $\beta>0$ is specified for SSNO.

\subsection*{Baseline methods}

We benchmark SSNO against state-of-the-art neural PDE solvers, including the Fourier-based neural operators DPOT~\cite{hao2024dpot} and FFNO~\cite{tran2023factorized}, the convolutional-based PDE solver CNext~\cite{liu2022convnet,ohana2024well}, the transformer-based PDE solver FactFormer~\cite{li2023scalable}, and the physics-encoded model PeRCNN~\cite{rao2023encoding}, which embeds FD stencils into its architecture. For 3D cases, the set of baseline models is consistent with the 2D setting, including DPOT, FFNO, CNext, and the physics-encoded PeRCNN. For the ACE on the sphere, we adopt Spherical FNO (SFNO)~\cite{bonev2023spherical} as a baseline, which generalizes FNO to spherical geometries by replacing the Fourier transformation with spherical harmonic convolutions.

\paragraph{DPOT.}
We adopt DPOT~\cite{hao2024dpot}, an operator transformer with auto-regressive denoising pre-training and Fourier attention, using the official codebase at \url{https://github.com/HaoZhongkai/DPOT}. We tune depth, width, spectral modes, learning rate, and rollout schedule. Specifically, we run a grid search over
layers $\in\{4,6\}$, width $\in\{32,64\}$, Fourier modes $\in\{12,16,32\}$, learning rate ($lr$) $\in\{10^{-3},5\times10^{-3},10^{-2}\}$, and rollout $n_1\in\{1,4\},\ n_2\in\{4,8\}$ for NSE and $n_1\in\{1,2\},\ n_2\in\{1,2,4\}$ for others,
and select the best configuration on the validation split (seed fixed to $0$). The best hyperparameters per PDE benchmark for DPOT are reported in Table~\ref{tab:dpot_best}.

\begin{table}[t!]
\centering
\caption{Best hyperparameters per PDE benchmark for DPOT.}
\label{tab:dpot_best}
\small
\begin{tabular}{lcccccc}
\toprule
\textbf{Dataset} & {Layers} & {Width} & {Modes} & $lr$ & $(n_1,n_2)$ & {\# Params (K)} \\
\midrule
2D KSE                 & 6 & 64 & 16 & $1\times10^{-2}$ & 1/2 & 227.0 \\
3D KSE                 & 4 & 64 & 16 & $5\times10^{-3}$ & 1/1 & 2212.6 \\
2D SHE                 & 6 & 64 & 16 & $1\times10^{-2}$ & 1/2 & 227.0 \\
3D SHE                 & 4 & 64 & 16 & $1\times10^{-2}$ & 1/1 & 2212.6 \\
2D NSE (Re = 500)                  & 6 & 64 & 32 & $1\times10^{-2}$ & 4/4 & 423.6 \\
2D NSE (Re = 1000)                 & 6 & 64 & 32 & $5\times10^{-3}$ & 1/4 & 423.6 \\
2D NSE (Re = 1500)                & 6 & 64 & 32 & $1\times10^{-2}$ & 4/4 & 423.6 \\
\bottomrule
\end{tabular}
\end{table}

\paragraph{FFNO.} We adopt FFNO~\cite{tran2023factorized}, which improves upon FNO by factorizing the Fourier kernel into separable components, which reduces parameter count and memory consumption while enabling deeper and more expressive networks. We use the implementation from \url{https://github.com/alasdairtran/fourierflow}. We tune depth, width, spectral modes, learning rate, and the rollout schedule. Specifically, we run a grid search over
layers $\in\{4,6\}$, width $\in\{16, 32\}$, Fourier modes $\in\{32, 48\}$ for NSEs and $\in\{24, 32\}$ for others, $lr$ $\in\{10^{-3},5\times10^{-3},10^{-2}\}$, and rollout $n_1\in\{1,4\},\ n_2\in\{4,8\}$ for NSE and $n_1\in\{1,2\},\ n_2\in\{1,2,4\}$ for others,
and select the best configuration on the validation split (seed fixed to $0$). The best hyperparameters per PDE benchmark for FFNO are reported in Table~\ref{tab:ffno_best}.

\begin{table}[t!]
\centering
\caption{Best hyperparameters per PDE benchmark for FFNO.}
\label{tab:ffno_best}
\small
\begin{tabular}{lcccccc}
\toprule
\textbf{Dataset} & {Layers} & {Width} & {Modes} & $lr$ & $(n_1,n_2)$ & {\# Params (K)} \\
\midrule
2D KSE                 & 6 & 32 & 24 & $1\times10^{-2}$ & 1/2 & 644.8 \\
3D KSE                 & 4 & 32 & 32 & $1\times10^{-2}$ & 1/2 & 824.6 \\
2D SHE                 & 4 & 32 & 24 & $1\times10^{-2}$ & 1/2 & 431.4 \\
3D SHE                 & 6 & 32 & 32 & $5\times10^{-3}$ & 2/1 & 1234.6 \\
2D NSE (Re = 500)       & 4 & 16 & 48 & $1\times10^{-3}$ & 4/4 & 207.6 \\
2D NSE (Re = 1000)       & 6 & 32 & 48 & $5\times10^{-3}$ & 1/4 & 1234.6 \\
2D NSE (Re = 1500)       & 6 & 32 & 48 & $1\times10^{-2}$ & 4/4 & 1234.6 \\
\bottomrule
\end{tabular}
\end{table}

\paragraph{CNext.}
We adopt CNext~\cite{liu2022convnet,ohana2024well}, a family of modernized convolutional networks that borrow design choices from Vision Transformers while retaining the efficiency of ConvNets. In the recent benchmark study of~\cite{ohana2024well}, CNext attained state-of-the-art results across diverse PDE learning tasks. We use the reference implementation from the benchmark codebase~\cite{ohana2024well}. We tune depth, width, learning rate, and the rollout schedule. Specifically, we run a grid search over
layers $\in\{3,4\}$, width $\in\{16,32\}$, $lr$ $\in\{10^{-3},5\times10^{-3},10^{-2}\}$, and rollout $n_1\in\{1,4\},\ n_2\in\{4,8\}$ for NSE and $n_1\in\{1,2\},\ n_2\in\{1,2,4\}$ for others. The best hyperparameters per PDE benchmark for CNext are reported in Table~\ref{tab:cnext_best}.

\begin{table}[t!]
\centering
\caption{Best hyperparameters per PDE benchmark for CNext.}
\label{tab:cnext_best}
\small
\begin{tabular}{lccccc}
\toprule
\textbf{Dataset} & {Layers} & {Width} & $lr$ & $(n_1,n_2)$ & {\# Params (K)} \\
\midrule
2D KSE                 & 3 & 16 & $1\times10^{-3}$ & 1/2 & 844.0 \\
3D KSE                 & 4 & 16 & $1\times10^{-2}$ & 1/2 & 4219.0 \\
2D SHE                 & 4 & 32 & $1\times10^{-2}$ & 1/1 & 12954.6 \\
3D SHE                 & 3 & 16 & $5\times10^{-3}$ & 1/2 & 1203.1 \\
2D NSE (Re = 500)        & 3 & 32 & $1\times10^{-2}$ & 4/4 & 3265.0 \\
2D NSE (Re = 1000)       & 3 & 16 & $1\times10^{-2}$ & 1/8 & 844.0 \\
2D NSE (Re = 1500)       & 4 & 16 & $1\times10^{-2}$ & 1/8 & 3296.7 \\
\bottomrule
\end{tabular}
\end{table}

\paragraph{FactFormer.}
We adopt FactFormer~\cite{li2023scalable}, a transformer-based PDE solver that uses an axial factorization of the kernel integral via a learnable projection operator, enabling efficient modeling in multi-dimensional settings. We use the implementation from \url{https://github.com/BaratiLab/FactFormer}. We tune depth, width, number of attention heads, learning rate, and the rollout schedule. Concretely, we run a grid search over
layers $\in\{4,6\}$, width $\in\{32,64\}$, heads $\in\{4,8\}$, $lr$ $\in\{10^{-3},5\times10^{-3},10^{-2}\}$, and rollout $n_1\in\{1,4\},\ n_2\in\{4,8\}$ for NSE and $n_1\in\{1,2\},\ n_2\in\{1,2,4\}$ for others. The best hyperparameters per PDE benchmark for FactFormer are reported in Table~\ref{tab:factformer_best}.

\begin{table}[t!]
\centering
\caption{Best hyperparameters per PDE benchmark for FactFormer.}
\label{tab:factformer_best}
\small
\begin{tabular}{lcccccc}
\toprule
\textbf{Dataset} & {Layers} & {Width} & {Heads} & $lr$ & $(n_1,n_2)$ & {\# Params (K)} \\
\midrule
2D KSE                 & 4 & 64 & 4 & $5\times10^{-3}$ & 1/4 & 903.3 \\
2D SHE                  & 4 & 64 & 4 & $5\times10^{-3}$ & 1/2 & 903.3  \\
2D NSE (Re = 500)        & 6 & 32 & 4 & $5\times10^{-3}$ & 1/8 & 322.2 \\
2D NSE (Re = 1000)       & 6 & 64 & 4 & $5\times10^{-3}$ & 1/4 & 1322.8 \\
2D NSE (Re = 1500)        & 6 & 64 & 4 & $5\times10^{-3}$ & 1/4 & 1322.8 \\
\bottomrule
\end{tabular}
\end{table}

\paragraph{PeRCNN (physics-encoded).}
We adopt PeRCNN~\cite{rao2023encoding}, which embeds finite-difference stencils into the network. We use the implementation from \url{https://github.com/Raocp/PeRCNN}. We tune the learning rate, channel width, kernel size, and the rollout schedule. Concretely, we run a grid search over width $\in\{8,16,32,64\}$, kernel size $\in\{1,3,5\}$, $lr\in\{10^{-3},5\times10^{-3},10^{-2}\}$, and rollout $n_1\in\{1,4\},\ n_2\in\{4,8\}$ for NSE and $n_1\in\{1,2\},\ n_2\in\{1,2,4\}$ for others. The best hyperparameters per PDE benchmark for PeRCNN are reported in Table~\ref{tab:percnn_best}.

\begin{table}[t!]
\centering
\caption{Best hyperparameters per PDE benchmark for PeRCNN.}
\label{tab:percnn_best}
\small
\begin{tabular}{lccccc}
\toprule
\textbf{Dataset} & {Width} & {Kernel size} & $lr$ & $(n_1,n_2)$ & {\# Params (K)} \\
\midrule
2D KSE                 & 64 & 5 & $1\times10^{-2}$ & 2/2 & 5.0 \\
3D KSE                 & 16 & 5 & $1\times10^{-2}$ & 1/2 & 8.2 \\
2D SHE                  & 16 & 5 & $1\times10^{-2}$ & 2/2 & 1.3 \\
3D SHE                 & 32 & 5 & $1\times10^{-2}$ & 1/2 & 16.2 \\
2D NSE (Re = 500)       & 16 & 5 & $1\times10^{-3}$ & 4/8 & 1.3 \\
2D NSE (Re = 1000)      & 64 & 5 & $1\times10^{-3}$ & 4/8 & 5.0 \\
2D NSE (Re = 1500)       & 64 & 1 & $1\times10^{-2}$ & 4/8 & 0.3 \\
\bottomrule
\end{tabular}
\end{table}

\paragraph{SFNO (sphere).}
We adopt SFNO~\cite{bonev2023spherical}, which extends FNO to spherical geometries by replacing planar Fourier transforms with spherical-harmonic convolutions. We use the implementation from \url{https://github.com/NVIDIA/torch-harmonics}. We tune depth, width, spherical modes, learning rate, and the rollout schedule. Concretely, we run a grid search over layers fixed $\in\{4,6\}$, width $\,\in\{16,32\}$, $lr\,\in\{10^{-3},5\times10^{-3},10^{-2},2\times10^{-2}\}$, and rollout $n_1\in\{1,2\},\ n_2\in\{1,2,4\}$. The best hyperparameters per PDE benchmark for SFNO are reported in Table~\ref{tab:sfno_best}.

\begin{table}[t!]
\centering
\caption{Best hyperparameters per PDE benchmark for SFNO.}
\label{tab:sfno_best}
\small
\begin{tabular}{lccccc}
\toprule
\textbf{Dataset} & {Layers} & {Width} &  $lr$ & $(n_1,n_2)$ & {\# Params (K)} \\
\midrule
ACE on $\mathbb{S}^2$ & 4 & 16 &  $2\times10^{-2}$ & 1/1 & 90.3 \\
\bottomrule
\end{tabular}
\end{table}

\paragraph{SSNO.}
We vary the number of spectral channels, anti-aliasing, and the temporal unroll. The hyperparameter grid is:
$k_{\text{num}}\in\{8,16\}$, \texttt{use\_fu}$\in\{0,1\}$, \texttt{use\_pi}$=\{0,1\}$, learning rate $lr\in\{10^{-3},5\times10^{-3},10^{-2}\}$, and rollout $n_1\in\{1,4\},\ n_2\in\{4,8\}$ for NSE and $n_1\in\{1,2\},\ n_2\in\{1,2,4\}$ for others. Here $k_{\text{num}}$ is the spectral channel count used by the SSNO operator and equals the output dimension of the Freq2Vec embedding. The switch \texttt{use\_fu} enables the reaction block and \texttt{use\_pi} toggles the $\Pi$ block. The best hyperparameters per PDE benchmark for SSNO are reported in Appendix Table~\ref{tab:ssno_best}.

\begin{table}[t!]
\centering
\caption{Best hyperparameters per PDE benchmark for SSNO.}
\label{tab:ssno_best}
\small
\begin{tabular}{lcccccc}
\toprule
\textbf{Dataset} & $k_{\text{num}}$ & \texttt{use\_fu} & \texttt{use\_pi} & $lr$ & $(n_1,n_2)$& {\# Params (K)} \\
\midrule
2D KSE                 & 8 & 1 & 1 & $1\times10^{-2}$ & 1/2 & 15.7 \\
3D KSE                 & 8 & 1 & 1 & $1\times10^{-2}$ & 1/2  & 15.8 \\
2D SHE                 & -  & 1 & 0 & $1\times10^{-2}$ & 1/2 & 8.9 \\
3D SHE                 & 16 & 1 & 1 & $1\times10^{-2}$ & 1/1 & 17.9 \\
2D NSE (Re = 500)        & 16 & 0 & 1 & $1\times10^{-2}$ & 4/8  & 17.9 \\
2D NSE (Re = 1000)       & 8 & 1 & 0 & $1\times10^{-2}$ & 4/8  & 15.8 \\
2D NSE (Re = 1500)       & 8  & 1 & 1 & $1\times10^{-2}$ & 1/8 & 20.1 \\
ACE on $\mathbb{S}^2$  & -  & 1 & 0 & $5\times10^{-3}$ & 2/4 & 8.8 \\
\bottomrule
\end{tabular}
\end{table}

\section*{Supplementary Note C: Snapshots of evolution trajectories}
\begin{figure}[t]
    \centering
    \includegraphics[width=0.9\linewidth]{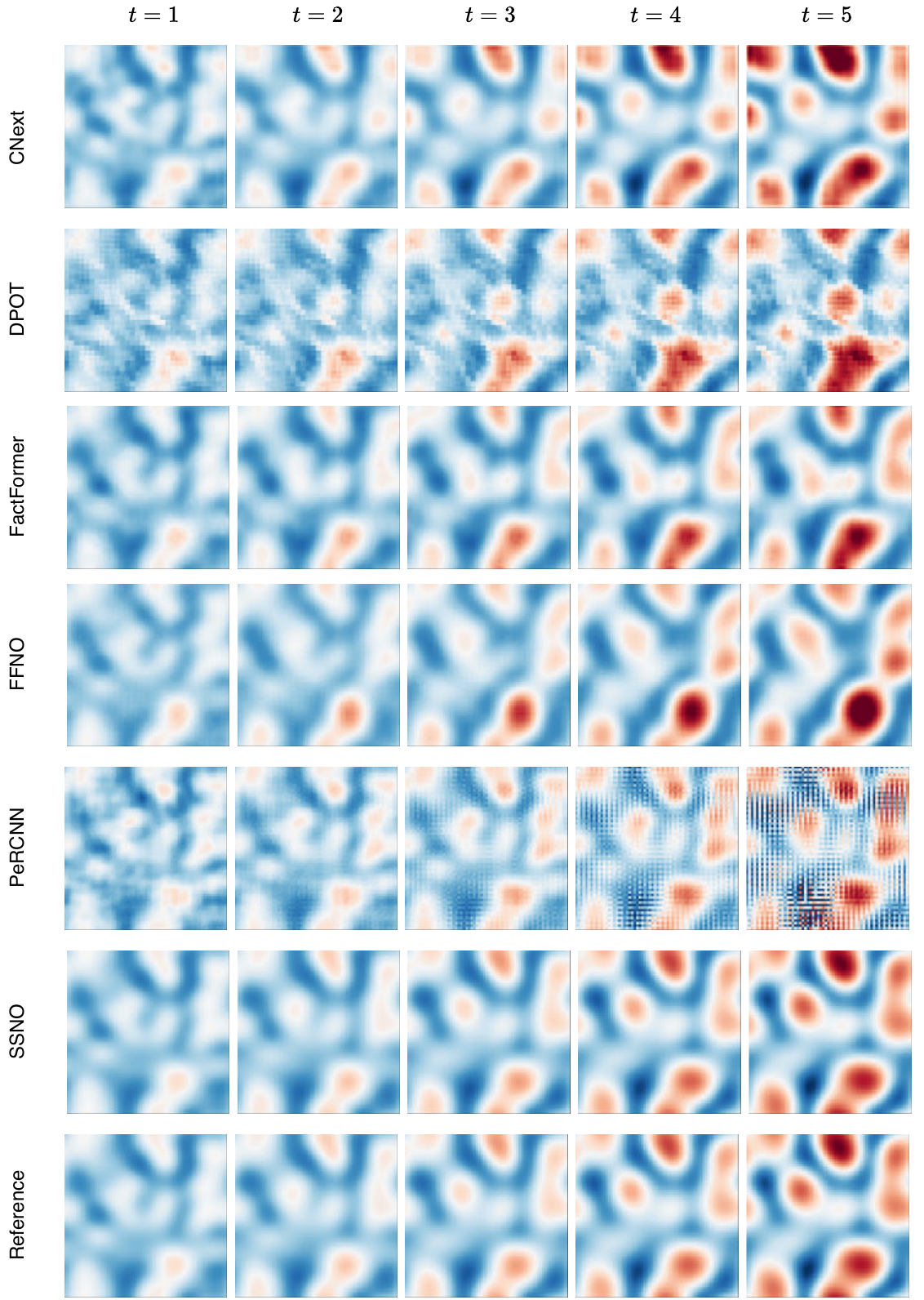}
    \caption{\textbf{2D KSE: snapshots of evolution trajectories.} Rollouts on the 2D Kuramoto--Sivashinsky equation for SSNO, the baselines, and the reference.}
    \label{appfig1}
\end{figure}

\begin{figure}[t]
    \centering
    \includegraphics[width=0.9\linewidth]{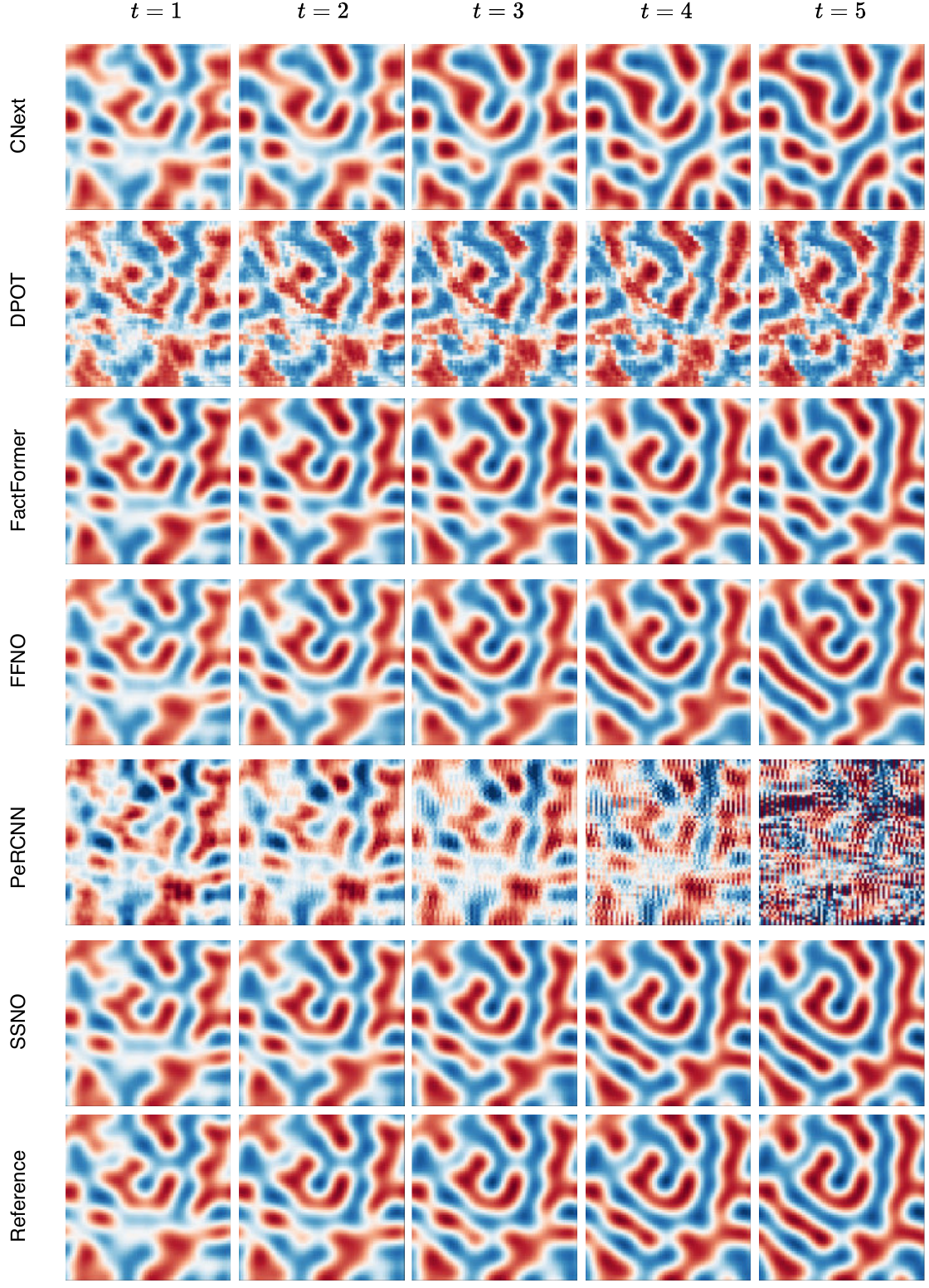}
    \caption{\textbf{2D SHE: snapshots of evolution trajectories.} Rollouts on the 2D Swift--Hohenberg equation for SSNO, the baselines, and the reference.}
    \label{appfig2}
\end{figure}

\begin{figure}[t]
    \centering
    \includegraphics[width=0.9\linewidth]{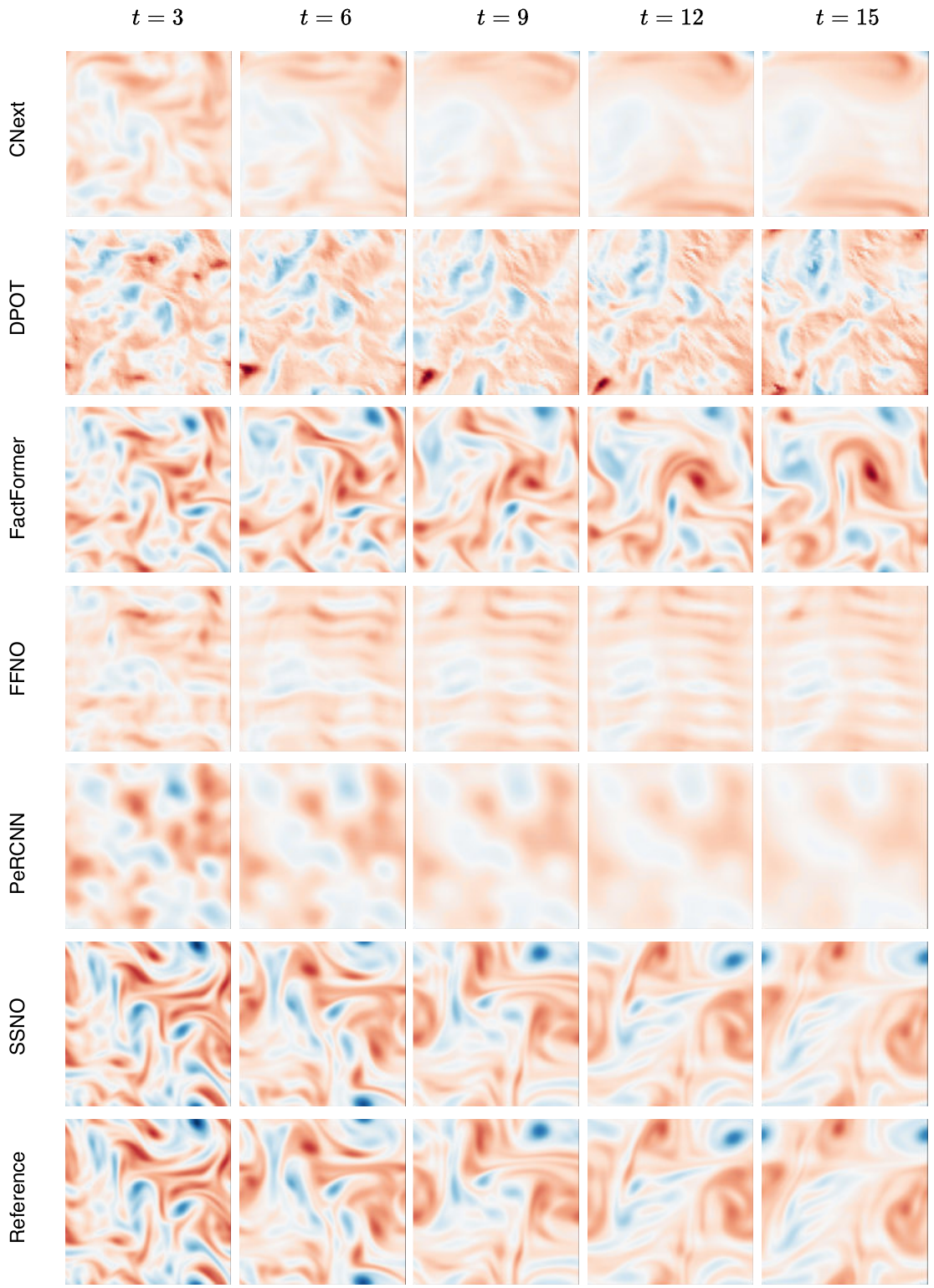}
    \caption{\textbf{2D NSE (Re = 500): snapshots of evolution trajectories.} Rollouts on the 2D incompressible Navier--Stokes system at $\mathrm{Re}{=}500$ for SSNO, the baselines, and the reference.}
    \label{appfig3}
\end{figure}

\begin{figure}[t]
    \centering
    \includegraphics[width=0.9\linewidth]{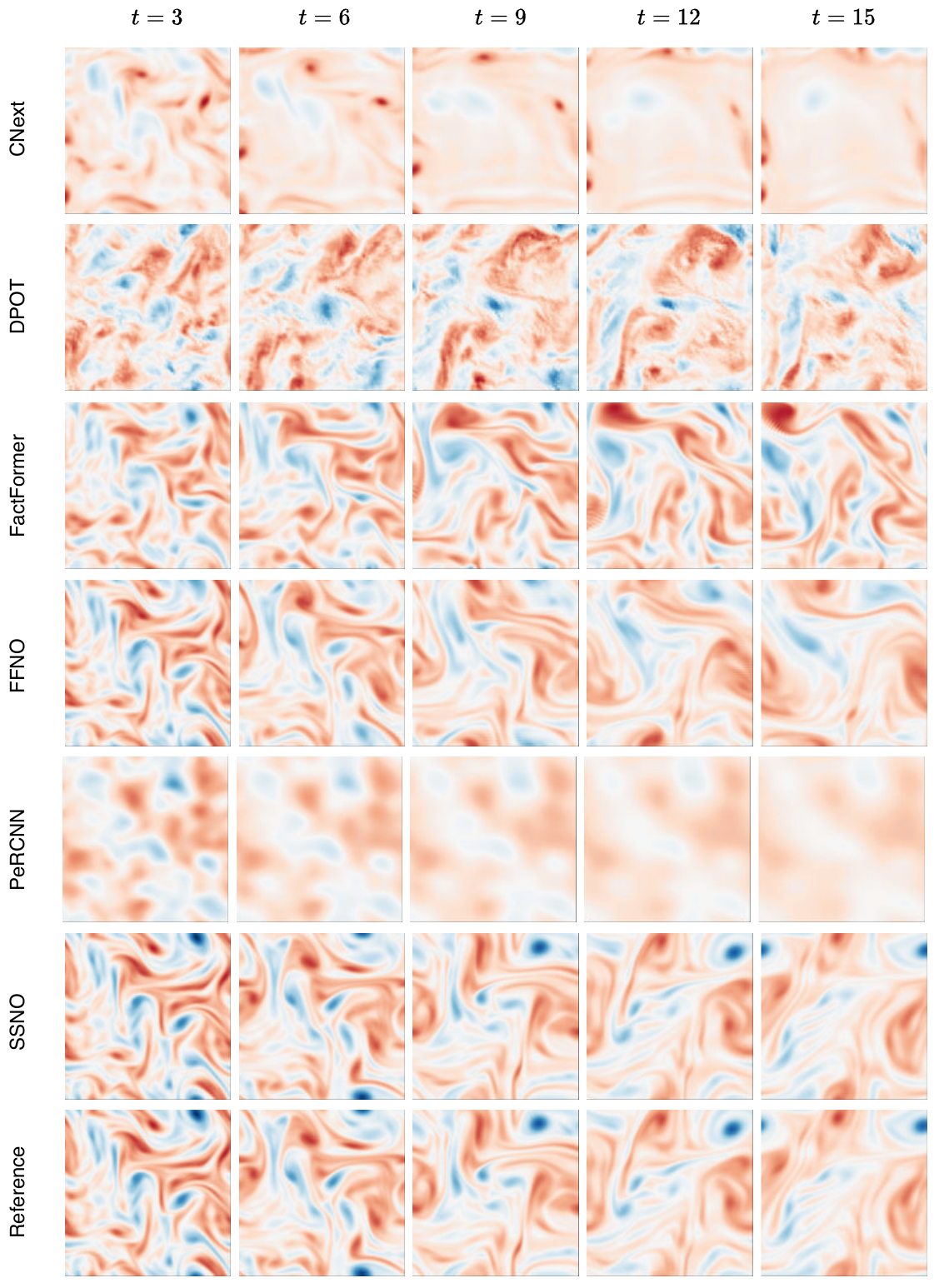}
    \caption{\textbf{2D NSE (Re = 1000): snapshots of evolution trajectories.} Rollouts on the 2D incompressible Navier--Stokes system at $\mathrm{Re}{=}1000$ for SSNO, the baselines, and the reference.}
    \label{appfig4}
\end{figure}

\begin{figure}[t]
    \centering
    \includegraphics[width=0.9\linewidth]{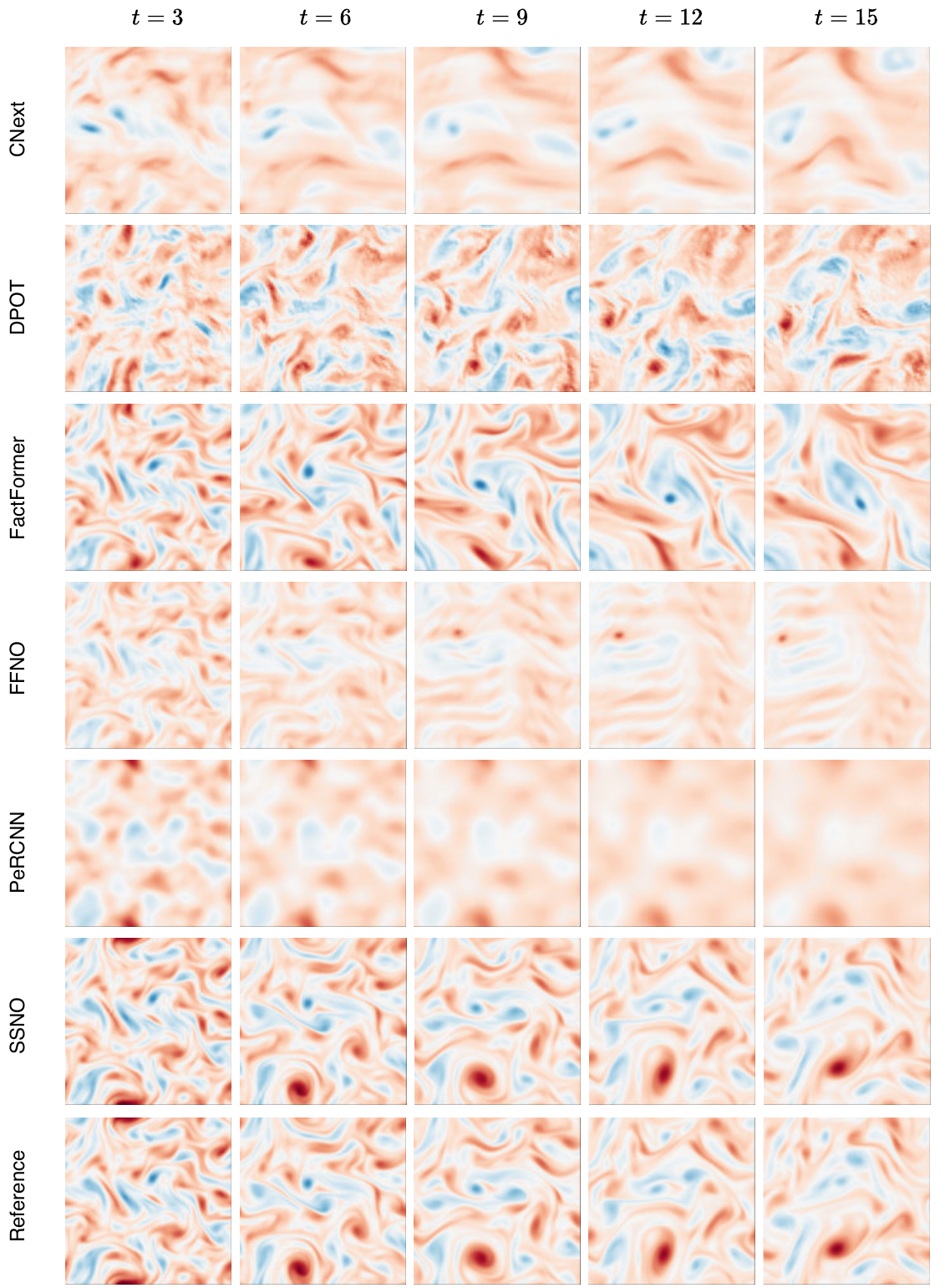}
    \caption{\textbf{2D NSE (Re = 1500): snapshots of evolution trajectories.} Rollouts on the 2D incompressible Navier--Stokes system at $\mathrm{Re}{=}1500$ for SSNO, the baselines, and the reference.}
    \label{appfig5}
\end{figure}

\begin{figure}[t]
    \centering
    \includegraphics[width=0.9\linewidth]{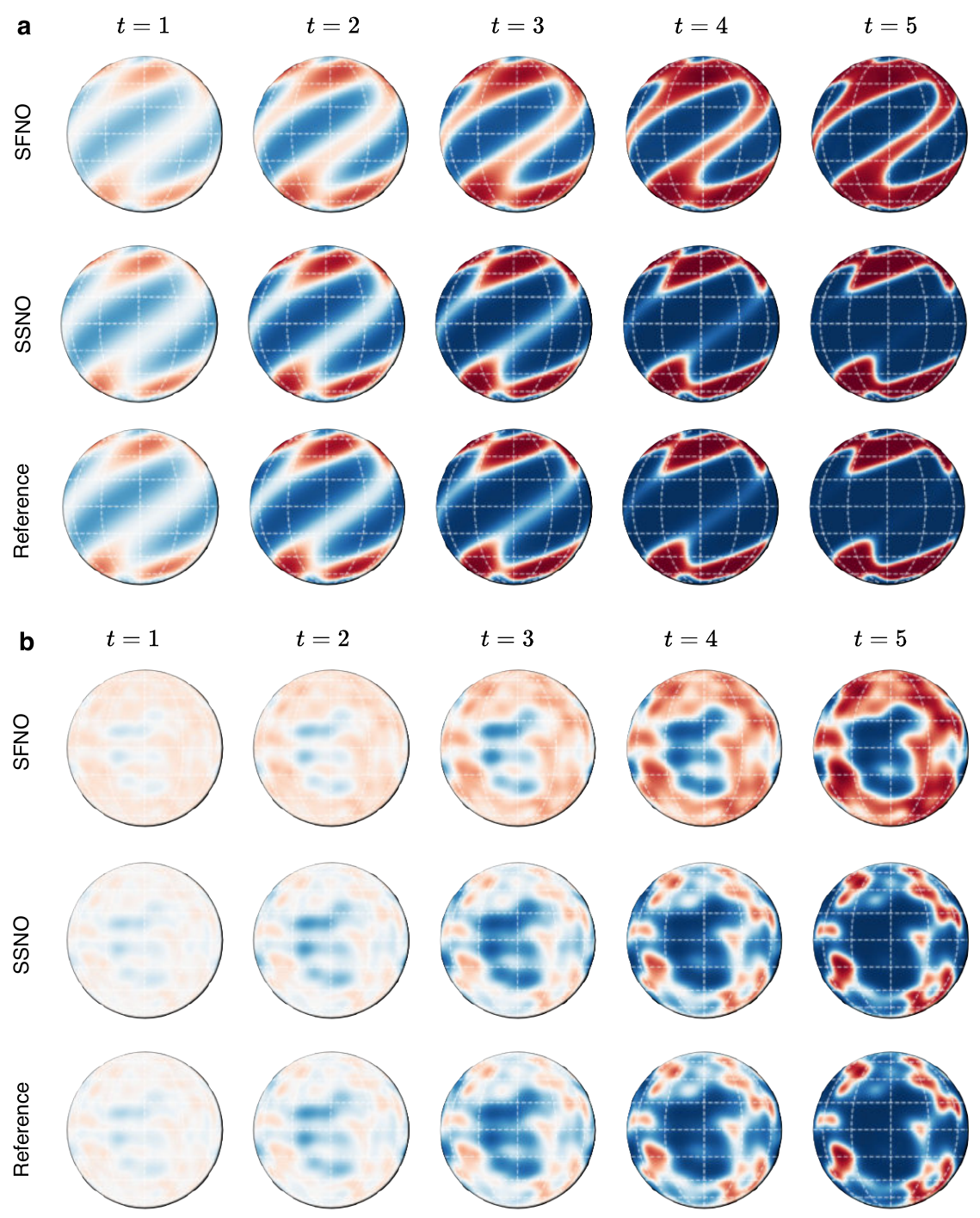}
    \caption{\textbf{ACE on $\mathbb{S}^2$: snapshots of evolution trajectories.} Rollouts on the Allen--Cahn equation on the sphere for SSNO, the baselines, and the reference. \textbf{a}. OOD case with IC1. \textbf{b}. OOD case with IC3.}
    \label{appfig6}
\end{figure}

\begin{figure}[t]
    \centering
    \includegraphics[width=0.9\linewidth]{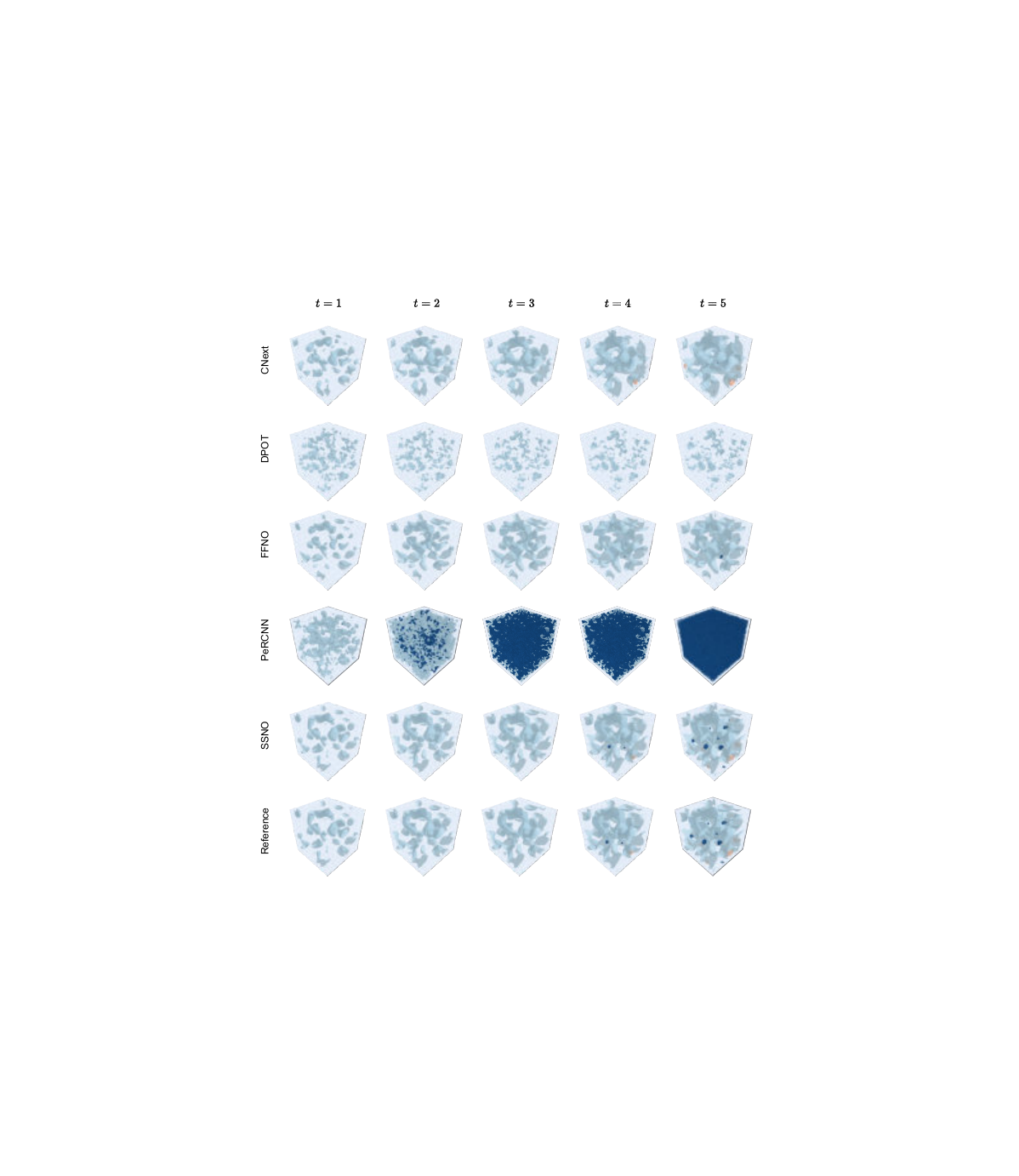}
    \caption{\textbf{3D KSE: snapshots of evolution trajectories.} Rollouts on the 3D Kuramoto--Sivashinsky equation for SSNO, the baselines, and the reference.}
    \label{appfig7}
\end{figure}

\begin{figure}[t]
    \centering
    \includegraphics[width=0.9\linewidth]{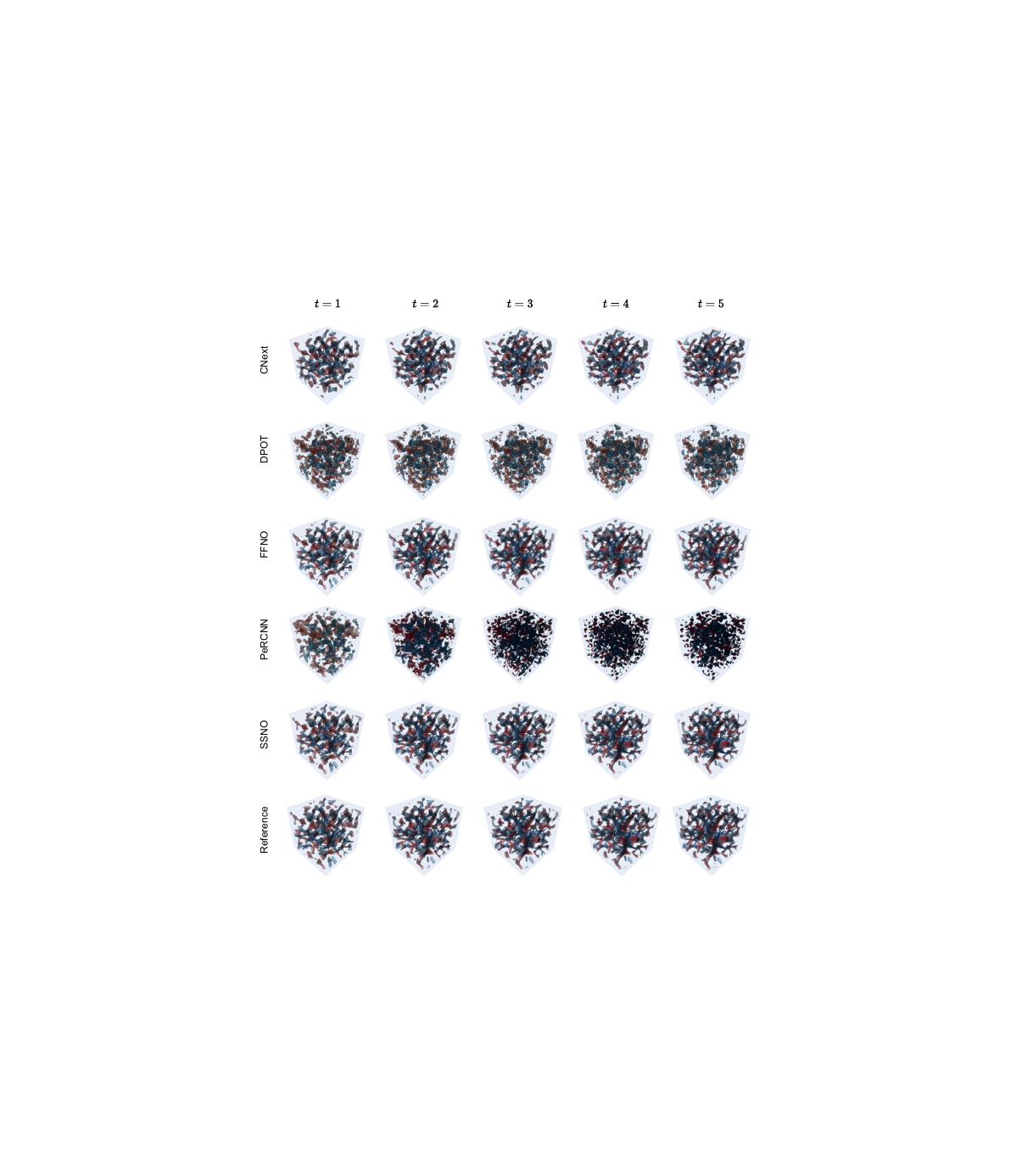}
    \caption{\textbf{3D SHE: snapshots of evolution trajectories.} Rollouts on the 3D Swift--Hohenberg equation for SSNO, the baselines, and the reference.}
    \label{appfig8}
\end{figure}

We visualize end-to-end rollouts for each benchmark to compare SSNO against the baselines and the ground-truth solution: 2D KSE (Appendix Fig.~\ref{appfig1}), 2D SHE (Appendix Fig.~\ref{appfig2}), 2D NSE at $\mathrm{Re}=500$ (Appendix Fig.~\ref{appfig3}), $\mathrm{Re}=1000$ (Appendix Fig.~\ref{appfig4}), and $\mathrm{Re}=1500$ (Appendix Fig.~\ref{appfig5}), ACE on $\mathbb{S}^2$ (Appendix Fig.~\ref{appfig6}), 3D KSE (Appendix Fig.~\ref{appfig7}), and 3D SHE (Appendix Fig.~\ref{appfig8}).

\end{document}